\documentclass[aps,reprint,nofootinbib]{revtex4-1}
\usepackage{graphicx,amsfonts,amsmath,amssymb}

\usepackage{bm}
\usepackage{url}
\makeatletter
\g@addto@macro{\UrlBreaks}{\UrlOrds}
\makeatother

\begin{document}
\title{The Vertical Logic of Hamiltonian Methods (Part 1)}
\date{\today}
\author{C. Baumgarten}

\email{christian-baumgarten@gmx.net}

\def\begeq{\begin{equation}}
\def\endeq{\end{equation}}
\def\bquo{\begin{quotation}}
\def\equo{\end{quotation}}
\def\begary{\begeq\begin{array}}
\def\endary{\end{array}\endeq}
\newcommand{\myarray}[1]{\begin{equation}\begin{split}#1\end{split}\end{equation}}
\def\bmtx{\left(\begin{array}}
\def\emtx{\end{array}\right)}
\def\d{\partial}
\def\h{\eta}
\def\w{\omega}
\def\W{\Omega}
\def\s{\sigma}
\def\eps{\varepsilon}
\def\e{\epsilon}
\def\a{\alpha}
\def\b{\beta}
\def\g{\gamma}
\def\G{\Gamma}
\def\y{\gamma}
\def\d{\partial}
\def\S{{\Sigma}}

\def\leftD#1{\overset{\leftarrow}{#1}}
\def\rightD#1{\overset{\rightarrow}{#1}}

\begin{abstract}
We discuss the key role that Hamiltonian notions play in physics.
Five examples are given that illustrate the versatility and generality of 
Hamiltonian notions. The given examples concern the interconnection between
quantum mechanics, special relativity and electromagnetism. 
It will be demonstrated that a derivation of these core concepts of
modern physics requires little more than a an abstract classical 
analysis of linear Hamiltonian theory. 
\end{abstract}


\maketitle

\section{Introduction}

It is a tradition to celebrate that physics is able to provide
us with the deepest possible insights into the nature of reality.
And it is true, of course -- this is exactly what physics does.
There is a tendency to think that, for providing these deep insights,
physics must be in the possession of the knowledge of deep and 
profound principles. We shall provide examples that support
a different view, namely that not everything, which is considered
to be a deep physical insight, requires an explanation that stems
from deep and profound principles. Deep and profound are not so much
the principles but the conclusions that bright people were and still
are able to draw from them.

This article has two parts. In this first part we provide some examples that 
illustrate, as we think, how essential concepts of modern physics can be 
developed 'naturally' on the basis of classical Hamiltonian methods. 
In part two (forthcoming), we shall discuss these findings in more depth 
and provide an outlook to further possible developments.

Hamiltonian notions are usually introduced to students as a
reformulation of Newtonian mechanics.
However, as Whittaker reported in his famous treatment on analytical
dynamics\footnote{See page 265 in Ref.~\cite{Whittaker}.}, a theorem
due to Lie and Koenigs has demonstrated that ``all the
differential equations which arise from problems in the Calculus of
Variations, with one independent variable, can be expressed in the
Hamiltonian form.'' 

If this result is correct, then Hamiltonian methods are mostly
mathematical and do not derive from the axioms of Newtonian mechanics
nor from any other metaphysical principle\footnote{
For a discussion of the metaphysical problems related to the 
{\it principle of least action}, see for instance Ref.~\cite{Terekhovich}.
}. Instead this theorem suggest that the Hamiltonian form has a
general significance for any {\it thinkable} form of classical
mechanics. At least to the degree to which classical mechanics is
the physics of  ``differential equations which arise from problems in
the Calculus of Variations, with one independent variable''.

Even if it might perhaps be a historical truth that Hamiltons
equations of motion were found as a reformulation of Newtonian
mechanics, mediated by Lagrangian mechanics, the theorem of Lie
and Koenigs demonstrates that Hamiltonian methods are more general.
They can be applied in all branches of science in which dynamical
systems are of interest. There are applications in
thermodynamics~\cite{Baldiotti,Rajeev}, but also in
biology~\cite{Kerner1971,Paine} and epidemiology~\cite{BALLESTEROS2020}.
Strictly speaking, Newtonian mechanics has been falsified by Einstein's
special theory of relativity (STR). Though it remains a valid approximation
for small velocities, the conceptual basis of special relativity is
fundamentally different from that of Newtonian physics. Every student
learns ``Newton's axioms'', but Newton himself called them {\it definitiones}.
But ``[...] the traditional meaning of axiom [is] a self-evident first
proposition, which is neither provable nor in need of a proof''~\cite{Pulte}.
Hence we think that Newton's attitude towards time, space and motion are the
real axioms of Newtonian mechanics, because these were the elements
of his theory that Newton really considered to be self-evident~\cite{NewtonPrincipiae}:
\bquo
I do not define time, space, place and motion, as being well known to all.
\equo
Hamiltonian methods need nothing but the raw ingredients of a general
dynamical system: dynamical variables, an independent variable (mostly ``time'') and a constant of motion. They can be used to describe classical mechanics,
but relativistic mechanics as well.

Newtons ``self-evident'' understanding of space and time,
the {\it common sense interpretation}, turned out to be wrong.
But also his {\it definitions}, which are presented to students
as {\it axioms} or as {\it laws}, are not unproblematic.
Henri Poincare, for instance, remarked~\cite{Poincare}:
\bquo
{\it The Principle of Inertia} -- A body under the action
of no force can only move uniformly in a straight line.
Is this a truth imposed on the mind a priori? If this
be so, how is it that the Greeks ignored it? How could
they have believed that motion ceases with the cause of
motion? or, again, that every body, if there is nothing to
prevent it, will move in a circle, the noblest of all forms
of motion?
If it be said that the velocity of a body cannot change,
if there is no reason for it to change, may we not just as
legitimately maintain that the position of a body cannot
change, or that the curvature of its path cannot change,
without the agency of an external cause? Is, then, the
principle of inertia, which is not an a priori truth, an
experimental fact? Have there ever been experiments on
bodies acted on by no forces? and, if so, how did we know
that no forces were acting?
\equo
The status of Newtons ``laws of motion'' is far from evident.
K.R. Symon described this as follows~\cite{Symon}:
\bquo
The status of Newton's first two laws, [...], is often the subject
of dispute. We may regard Eqs. (1.9) [$F=m\,a$] as defining force in
terms of mass and acceleration. In this case, Newton's first two laws
are not laws at all but merely definitions of a new concept to be
introduced in the theory.
\equo
Newton considered absolute space, absolute time and motion of
matter in space as fundamental and self-evident notions.
Einstein's STR tells us that Newton was wrong in this respect.
While his mechanics is still ``valid'' as an approximation for
small velocities, the ``absolute'' notions of space and time
can hardly be regarded as approximations of the ``relative''
notions that space and time became with the introduction of STR.
One might say that the law of inertia apparently remained fully
valid. But this ``law'' remains mysterious. It seems to claim a
cause not for change, but for a resistance against change.
Is such a ``law'' required?
And what is it's status? Is it a definition, an axiom or an
experimental fact about nature?
Poincare wrote~\cite{Poincare}:
\bquo
Has this generalised law of inertia been verified by
experiment, and can it be so verified? When Newton
wrote the Principia, he certainly regarded this truth as
experimentally acquired and demonstrated. It was so in
his eyes, not only from the anthropomorphic conception
to which I shall later refer, but also because of the work
of Galileo. It was so proved by the laws of Kepler. According to
those laws, in fact, the path of a planet is entirely determined
by its initial position and initial velocity; this, indeed, is what
our generalised law of inertia requires.
\equo
According to Poincare, the essence of the law of inertia is
that in mechanics, {\it two} initial values (per degree of freedom)
are required to fully determine the path of objects.
This is an intriguing view and a view that enters a new level of
abstraction. It is a Lagrangian or Hamiltonian point of view.

Newtonian mechanics, though it introduced a new level of abstraction
into the natural sciences, still suffered from a lack of abstraction.
The notions of Newtonian mechanics are too abstract to be intuitively
clear but in some sense they are not abstract enough.
Newtonian mechanics presupposed specific notions like space, time and
motion of massive objects as self-evident, as if these notions would
somehow be the basis of any {\it thinkable} physical world.

Hamiltonian mechanics presupposes nothing like that.
While Lagrangian mechanics derives from the beautiful but opaque
``principle of least action'', Hamiltonian notions can be derived
without any {\it specific} assumption, purely from the distinction
between those physical quantities that may vary in time from those
quantities that may not.
The former ones are called {\it dynamical variables},
the latter are called constants of motion (COMs), which are the conserved
quantities\footnote{It is part of the results of this article to show that 
also invariants like the rest mass can, on a basic level of 
Hamiltonian description, be described as constants of motion.}.
In preceding articles we argued on the basis of {\it pure} Hamiltonian 
theory, which follows from the assumption that {\it any} constant 
physical quantities with ontological content essentially derives from
constants of motion~\cite{qed_paper,osc_paper}.

The examples to be discussed will illustrate that the formal constraints 
which are imposed on dynamical equations by Hamiltonian notions, suffice
to derive the core elements of modern physics.

The first example (Sec.~\ref{sec_ex1}) is taken from accelerator physics 
and concerns an application of the Hamiltonian method which documents the 
remarkable fact, that Hamiltonian notions emerging from apparently 
disconnected levels of physical description fit together seamlessly. 
We call these interconnections of different levels ``vertical'' and shall
illustrate this in what follows\footnote{In part two of this article we shall 
try to expose the idea of vertical connections in part two in more 
mathematical detail. In this first part we proceed more intuitively.}.

In the second example (Sec.~\ref{sec_ex2}) we summarize and discuss
a two-page derivation of Schr\"odinger's equation from a simple 
classical Hamiltonian constraint on the dispersion relation~\cite{seq_paper}. 
In the third example (Sec.~\ref{sec_ex3}) we shall show that special relativity 
and the Dirac equation can be obtained from  pure Hamiltonian concepts.
The last two sections are dedicated to show how the Lorentz transformations
(Sec.~\ref{sec_ex4}) and finally Maxwell's equations (Sec.~\ref{sec_ex5})
fit into (or even follow from) the sketched Hamiltonian (symplectic) framework.

\section{First Example (Setup): Cyclotron Motion}
\label{sec_ex1}

\begin{figure}[t]
\parbox{8.5cm}{
\includegraphics[width=8.5cm,keepaspectratio]{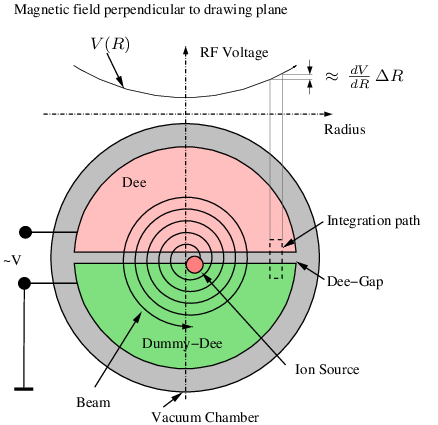}
\caption[Cyclotron]{
Classical cyclotron. Particles are accelerated when passing the gap
between the Dee and the (grounded) dummy-Dee. An example for a radius 
dependent gap voltage $V(R)$ is plotted above the sketched cyclotron. 
The integration path of Eq.~\ref{eq_rotE} is shown as a dashed rectangle.
\label{fig_cyclotron}
}}
\end{figure}

Many years ago a (now retired) colleague published a paper titled
``Application of the Phase Compression - Phase
Expansion Effect for Isochronous Storage Rings''~\cite{Joho}.
This is a very specialized topic, but the point we intend to make does not
require deep expertise in accelerators.
Consider a classical cyclotron (Fig.~\ref{fig_cyclotron}), i.e. particles in 
almost circular motion in a plane perpendicular to a homogeneous magnetic 
field $B$. It is well known that the motion of particles in electromagnetic
fields can be derived from the classical Hamiltonian function of a point particle.
This does neither exhausts the possibilities of the Hamiltonian method nor
does is solve the full problem. The solution of the equations of motion
obtained from the Hamiltonian of a relativistic particle in external
electromagnetic fields and yields some trajectory $(\vec x(t),\vec v(t))$, 
which is then used as a reference.

In order to verify the stability of a beam on the reference trajectory, 
accelerator physicists use the Hamiltonian techniques again to analyze the
{\it local} behavior for all starting conditions in the {\it vicinity} of
the reference trajectory, i.e. at $\vec x(0)+\delta x(t)$ and with velocity
$\vec v(0)+\delta\vec v(0)$. A real particle beam will only show up  
if a non-vanishing vicinity of starting conditions yields trajectories 
that stay in the vicinity of the reference trajectory, i.e. if the ``betatron 
motion'' is oscillatory\footnote{The orbits are stable if 
the ``tunes'' are real. In circular accelerators, the periodicity of 
distortions requires that more conditions have to be met to preserve
stability, typically that the tunes are not integers and not in some 
integer relation~\cite{Guignard}.}.
This is analyzed in the local co-moving frame, i.e. by solving the
equations of motion for $(\delta\vec x(t),\delta\vec v(t))$.

The importance of this type of stability analysis in mechanics depends, of
course, on the specific problem at hand. For the prediction of planetary
orbits of the next century, stability analysis can almost always be omitted.
But the reason to allow for this omission does not lie in the stability
of planetary orbits, but in the small number of revolutions relevant to us.
However, with respect to the long-term stability of a dynamical system,
stability analysis is inevitably required. Optical transitions in atoms
have frequencies in the range of a few hundred ${\rm THz}$. Nonetheless
most atoms\footnote{Those with a stable nucleus.} are stable. This stability
is, from a point-mechanical perspective, difficult to explain.

Coming back to the example: The circulation frequency $\w_c$ of a coasting
particle in a cyclotron is given by the ratio of the particle's velocity 
to the length of the closed orbit. It can be fine-tuned by the value of 
the magnetic field at the respective radius.
In so-called ``isochronous'' machines, the field is tuned, for instance
by iron shims or by trim-coils, in order to precisely synchronize the phase
$\phi$ between the radio frequency (rf) oscillation of the rf-electrodes,
in cyclotrons traditionally called ``Dees'', and the particle's
circulation frequency~\cite{Shimm}. 
During the passage of the acceleration gap, the circulating particles may
gain or lose energy, depending on the rf-phase $\phi$ at passage with
$\cos{\phi}$. The maximal energy gain $dE/dn$ for one turn can hence be
written as
\begeq
{dE\over dn}=E_G\,\cos{\phi}\,,
\label{eq_egain}
\endeq
where $E$ is the particle's kinetic energy, $E_G$ the maximal energy gain
per turn and $n$ is the turn number\footnote{
The use of the (discrete) turn number as a continuous independent 
variable is called the ``smooth acceleration approximation''.}. 
Eq.~\ref{eq_egain} fixes the phase of maximal energy gain to be zero. 
Note that in cyclotrons, the radius $R$ of the reference orbit rises 
monotonic with the kinetic energy $E$ so that one may express, for
particles in the vicinity of the reference orbit, radius by energy and 
vice versa. Since the phase is proportional to a time variable 
$\phi=\Delta\w\,t$, it is the Hamiltonian conjugate of energy.

The existence of a canonical pair $(E,\phi)$, formally proven by
Lie and Koenigs, allows to infer that Eq.~\ref{eq_egain} is one of
two Hamiltonian equations of motion and that a Hamiltonian function
${\cal H}(E,\phi)$ should exist such that Hamilton's equations of
motion hold true:
\myarray{
{dE\over dn}&={\d{\cal H}\over\d\phi}\\
{d\phi\over dn}&=-{\d{\cal H}\over\d E}\,.
\label{eq_cycH}
}
The combination of Eq.~\ref{eq_egain} and Eq.~\ref{eq_cycH} results in
\begeq
{\d{\cal H}\over\d\phi}=E_G\,\cos{\phi}\,,
\endeq
the integration of which then yields
\begeq
{\cal H}=E_G\,\sin{\phi}+F(E)\,,
\label{eq_Hcyc}
\endeq
where $F(E)$ is an integration ``constant''\footnote{
  $F(E)$ does not depend on $\phi$ and is hence a constant with
  respect to partial differentiation.} and describes 
the radial phase shift by the radial profile of the (static) magnetic
field as we described before. Hence $F(E)=0$ holds in a perfectly isochronous
machine.
However, inserting Eq.~\ref{eq_Hcyc} into the second of Eq.~\ref{eq_cycH} yields
another non-zero phase shift 
\begeq
{d\phi\over dn}=-{\d E_G\over\d E}\,\sin{\phi}\,.
\label{eq_phaseshift}
\endeq
which, because of it's phase dependency, must be related to the rf-acceleration.
The maximum energy gain $E_G$ is equal to the particle's charge multiplied
by the maximal Dee voltage $V(R)$ which may (but does not have to)
depend on radius (and hence on energy):  
\begeq
E_G=q\,V(R)
\endeq
Inserting this into Eq.~\ref{eq_phaseshift} results
\begeq
{d\phi\over dn}=-q\,{dV\over dR}\,{dR\over dE}\,\sin{\phi}\,.
\endeq
The term ${dR\over dE}$ is called radius gain with energy and
can be derived from $p=R\,q\,B$. Since $p$ rises monotonically
with energy, ${dR\over dE}$ is a monotonic relationsship.

Hence there is -- even in a perfectly isochronous magnetic field -- 
the possibility for a non-zero phase shift per turn if the
accelerating voltage depends on the radius. This dependency
can otherwise only be derived using Maxwell's equations~\cite{Joho}:
\myarray{
\d_t \int\vec B\,d\vec A&=-\int\,\vec\nabla\times\vec E\,d\vec A=-\oint\,\vec E\,\vec ds\\
B_{\rm rf}\,g&=-{dV\over dR}\,{\sin{(\phi)}\over\w_{\rm rf}}\,,
\label{eq_rotE}
}
where $g$ is the gap distance. Hence a voltage gradient is necessarily
accompanied by a non-zero magnetic high-frequency field.
The integration area (integration path) is shown in Fig.~\ref{fig_cyclotron}.
The electric field inside the Dees vanishes so that the right 
side of Eq.~\ref{eq_rotE} is proportional to the voltage 
difference between the corresponding radial positions 
when the integration passes the Dee gap.
The causal explanation is as follows: according to Maxwell's equations,
the gradient of the oscillating rf voltage is accompanied by an 
oscillating axial magnetic field, the average of which is 
proportional to $\sin{\phi}$ as seen by the particle along it's orbit. 
This rf contribution to the magnetic field {\it causes} a horizontal kick
that changes the orbit length and therefore results in a phase shift
\footnote{For further details of the calculation see Ref.~\cite{Joho}.
For 3D electromagnetic modeling see also Ref.~\cite{Schippers}.}.

\begin{figure}[t]
\parbox{8.5cm}{
\includegraphics[width=8.5cm,keepaspectratio]{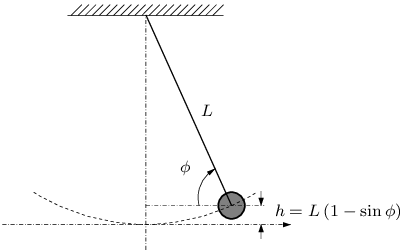}
\caption[Cyclotron]{
Classical pendulum. The potential energy is $m\,g\,h=m\,g\,L\,(1-\sin{\phi})$.
\label{fig_pendulum}
}}
\end{figure}

But how is it possible to derive the same result from a Hamiltonian 
that did not refer in any obvious way to the causal story? 
One might suspect that we ``somehow'' smuggled high-level knowledge 
(some content of Maxwell's equations) into EQ.~\ref{eq_Hcyc}. There is 
no doubt that a cyclotron is a device that can only be constructed with
sufficient knowledge of Maxwell's equations.
But EQ.~\ref{eq_Hcyc} is, up to a sign, identical to the momentum change
in a classical mechanical pendulum as depicted in Fig.~\ref{fig_pendulum}.
The equation of motion for the momentum is
\begeq
\dot p=-m\,g\,\cos{\phi}\,,
\endeq
which is (up to a sign) formally identical to EQ.~\ref{eq_Hcyc} for
the presumed condition $F(E)=0$. The only difference lies in the fact
that the mass $m$ and the gravitational constant $g$ do (``classically'') 
not vary with the momentum $p$. So if $\phi$ is 
(proportional to) the canonical conjugate of $p$, i.e. $x=\eps\phi$, 
then the presumed Hamiltonian equation
\begeq
\dot p=-{\d{\cal H}(p,\phi)\over\d(\eps\phi)}=-m\,g\,\cos{\phi}\,,
\endeq
leads to
\begeq
{\cal H}(x,p)=-\eps\,m\,g\,\sin{\phi}+T(p)+C
\endeq
and results in
\begeq
\dot x={\d{\cal H}(p,\phi)\over\d p}={dT\over dp}\,.
\endeq
and eventually with the correct value of the constants $\eps=L$
and $C=L\,m\,g$ to
\begeq
{\cal H}(x,p)=m\,g\,L\,(1-\sin{\phi})+T(p)=m\,g\,h+T(p)=\rm{const}
\endeq
It is hardly possible to defend the idea that an equation of motion of the
form of EQ.~\ref{eq_Hcyc} encodes high-level knowledge about
electromagnetism. If it would, why does the same equation describe a
simple pendulum where electrodynamics can safely be ignored?

The analytical (and predictive) power of the Hamiltonian method is
remarkable and far more versatile and general than Newtonian notions.
Nonetheless the physics curriculum, sticking to an historical account
of classical mechanics, teaches Hamiltonian mechanics as secondary,
as something which somehow derives from Newtonian mechanics. 

This has possibly it's legitimization in pedagogical reasons, but implies
that one derives a correct and powerful method, namely Hamiltonian mechanics,
not from basic {\it mathematical} principles, but instead from a falsified
and phenomenological theory, namely Newtonian mechanics.
Consequently every physics student has to un-learn (Newtonian based)
and re-learn (Hamiltonian based) physics in order to understand the
logic of quantum theory and (special) relativity.

Hamiltonian notions are used in various levels of description, often
treated separately in separate branches of physics. There is a Hamiltonian
for the Dirac particle, which is then sometimes replaced by a relativistic
point particle Hamiltonian, or by a Schr\"odinger Hamiltonian for an orbital
electron, then comes the Hamiltonian describing the inter-action that 
governs ionic binding in crystals, then some kind of classical 
Hamiltonian that describes the motion of the crystal being a grain of dust
in space and so on. In reality all these levels are interconnected, even
though they are treated in separate physics textbooks, books on quantum 
electrodynamics and quantum mechanics (QM), atomic physics, solid state 
physics and finally the grain of dust by classical mechanics or astrophysics.
There is no universally accepted coherent theoretical 
account known to the author describing the effects which physical
constraints resulting from one level of description have on other levels. 
From a birds eye view, physical reasoning is mostly horizontal, i.e. it stays
within one level. Little is known about the general patterns and the ``vertical''
interconnection between different levels of Hamiltonian description. 

The above example illustrates that we might in practice, when solving 
daily problems, for instance in accelerator physics, take a logical 
coherence of different levels for granted which is mostly ignored in
the usual presentation of physics.

\section{Second Example: Schr\"odinger's Equation}
\label{sec_ex2}

\begin{quotation}
The problem is not the interpretation of quantum mechanics. That's getting
things just backwards. The problem is the interpretation of classical
mechanics.\\
\mbox{}\hfill{\it -- Sidney Coleman}~\cite{Coleman}
\end{quotation}

In a preceding article we gave a short derivation of Schr\"odinger's 
equation~\cite{seq_paper}, which we shall briefly summarize and discuss in this
section. 

Assumed we would, for whatever reason, refuse the notion of  
the point particle and replace it with a (classical) density distribution 
$\rho(t,\vec x)$ in space and time. If the density is supposed to represent
a finite amount of matter, it must of course be normalizable:
\begeq
\int\,\rho(\vec x,t)\,d^3x=1\,.
\label{eq_density}
\endeq
The density must furthermore be positive semi-definite $\rho\ge 0$\footnote{
  These requirements are incidentally the same as one would demand for
  a probability distribution.} quantity.
This latter condition can be automatically fulfilled, if one expresses
$\rho$ by (possibly a sum of) squares of some auxiliary function
$\psi(\vec x,t)$.
Then $\psi$ is, due to Eq.~\ref{eq_density}, square-integrable
so that it's Fourier transform exists and is well-behaved.
The only reason and objective to use $\psi$ is to replace $\rho$ by a
positive semi-definite expression, so that we are free to opt for a complex
$\psi$, mostly for reasons of elegance and simplicity of the following
Fourier expressions. Hence we write the density as\footnote{
This choice obviously introduces a ``hidden variable'', namely the phase
of the complex function $\psi$, which can not be obtained by a measurement
of $\rho$.}  
\begeq
\rho=\psi^\dagger\psi\,.
\endeq
Then the Fourier transform can be written as
\begeq
\psi(t,\vec x)={1\over (2\,\pi)^2}\,\int\,\tilde\psi(\w,\vec k)\,\exp{[-i\,(\w\,t-\vec k\cdot\vec x)]}\,d^3k\,d\w\,.
\endeq
The translation of normalizability and positive definiteness into
square integrability directly leads to the space of $L^2$-normalizable
functions, i.e. to a Hilbert space. In a Hilbert space it is allowed to
freely select a basis which best suits to solve the problem. The use of
canonical transformations has the exact same purpose: To find a transformation
to new variables in which the solution is possible or - in the best case - 
trivial. The use of the Fourier transform is hence nothing but the choice of a
convenient basis. Since it is a reversible transformation, no information
can get lost by it's application.

Physics textbooks introduce wave packets as something alien to classical
physics which can only be legitimized empirically but not theoretically. 
It is part of this lore that the interference experiments with particles 
generate the surprising and counter-intuitive {\it necessity} to introduce 
a mechanical theory based on waves.
But this is not the only possible view and not the only possible motivation
to consider the use of wave packets in order to represent matter densities:
any {\it normalizable} semi-positive density can be represented by the square
of a complex wave function, i.e. by a ``wave-packet'', no matter if one
regards the idea to {\it use} a wave packet representation as (a consequence of)
an experimental discovery or merely as an arbitrary mathematical transformation.

The description of a density distribution by a wave packet is, as
we have just argued, little more than a free choice of basis in Hilbert
space. We might for instance be motivated by the fact that the notion of
the ``point particle'' is not an uncontroversial element of classical thought.
Maybe we just like to spell out the consequences of a purely mathematical
game. Or we might be driven by the question, why the transmission of physical
matter through space (i.e. ``motion'') should be described by different rules
of local causality compared to the transmission of energy by waves\footnote{This 
latter question receives full significance only in context of Einstein's 
theory of relativity, i.e. with the insight that also mass is a form 
of energy.}: Why would nature invent two types of motion? And if it would,
why then should they be related by a constant velocity of light, which builts the bridge
between relativistic mechanics (particle motion) and electrodynamics (i.e. wave motion)?

It is another math fact that, provided the frequencies $\w$ of the partial waves
are related to the wavelength by some relationship $\w=\w(\vec k)$, then
the ``velocity'' of the wave packet is in linear approximation given by the
group velocity\footnote{
The well-known linear expressions for the group velocity have first been
given by Sir W.R. Hamilton (sic!) in Ref.~\cite{Hamilton}.
See also Refs.~\cite{Brillouin,Whitham,Pain}.
}:
\begeq
\vec v_{gr}=\vec\nabla_k\,\w(\vec k)=({\d\w\over\d k_x},{\d\w\over\d k_y},{\d\w\over\d k_z})^T\,.
\label{eq_dispersion}
\endeq
As a matter of fact, the velocity of a classical particle
is given by an expression of the exact same mathematical form:
Hamilton's equations of motion (EQOM) for the coordinate velocity is 
\begeq
\dot q_i={\d{\cal H}\over\d p_i}\,.
\endeq
In vector notation, this reads
\begeq
\vec v=\vec\nabla_p\,{\cal H}(\vec p,\vec x)\,.
\label{eq_someHamiltonian}
\endeq
Hence the motion of the matter density $\rho$ can only be consistent with
(Hamiltonian) mechanics, if both velocities agree:
\begeq
\vec\nabla_{\vec k}\w(\vec k)=\vec\nabla_{\vec p}{\cal H}(\vec p,\vec x)\,,
\label{eq_correspondence}
\endeq
Since both descriptions are of full generality, Eq.~\ref{eq_correspondence}
must hold in any (Euklidean) space-time of any dimension.
The general equivalence of dispersion relations and Hamiltonians is
well-known, for instance in geometrical optics~\cite{Thorne2017}.
Hence this equivalence is just a math fact applicable to
any linear physical wave motion. It follows, by (partial) integration, that 
\myarray{
\vec k&\propto \vec p+\eps\,\vec A(\vec x)\\
\w(\vec k)&\propto {\cal H}(\vec p,\vec x)+\eps\,\phi(\vec x)\\
\label{eq_mech_cano}
}
where the additive ``constants'' $\phi$ and $\vec A$ are, in general, 
functions of the variables which are conjugate to the mechanical
momentum, i.e. the spatial position $\vec x$\footnote{
We shall come back to these quantities in the last example, in 
Sec.~\ref{sec_ex5}.}.

Therefore, given we use the conventional units of energy and frequency, 
some constant conversion factor must be introduced that allows to express 
energies in units of frequency and momenta in units of wavelength. 
This factor is usually represented by the symbol $\hbar$ and it's value
is solely determined by the historical choice of units.

Hermann Weyl explained in 1930 that the significance of the two constants
$\hbar$ and $c$ is, from a logical viewpoint, equivalent -- they are both
scaling constants which can not be obtained theoretically~\cite{Weyl1930}:
\bquo
The constants $c$ and $h$, the velocity of light and the quantum of action,
have caused some trouble. The insight into the significance of these
constants, obtained by the theory of relativity on the one hand and quantum
theory on the other, is most forcibly expressed by the fact that they do not
occur in the laws of Nature in a thoroughly systematic development of these
theories.
\equo
The impossibility to obtain the value of these constants theoretically is
due to the fact that both are the result of a contingent (historical)
choice of units.

The appearance of the action constant is often {\it interpreted} as if
it would provide evidence that energy, by means of some mysterious
logical leap, is ``chunked in portions''. But as we have just seen, the
appearance of an action constant can be derived from a small number
of mathematical assumptions: it is mostly the mathematical consequence 
of a continuity assumption, namely that matter should fill out space
continuously, which is quite opposite of ``chunking''. 
Eq.~\ref{eq_correspondence} reveils how and why the introduction of 
an universal action constant at some ``fundamental level'' leads to 
a kind of {\it wave-particle correspondence}: In order to obtain a 
wave-packet, the center of which moves with a {\it variable} velocity,
we need to equate both, the dispersion relation of the wave and
the velocity as defined by the Hamiltonian formalism. At first sight,
it does not seem to establish a ``duality'' of physical pictures.

We admit that the density $\rho=\psi^*\psi$ {\it can not} simply
be reinterpreted as a charge density in the sense of classical
electrodynamics. Firstly, there was no mention of a charge to this point. 
And secondly, the (self-) energy of a charge distribution would have 
to be directly related to the (spatial) size of the wave-packet.
The above reasoning, however, suggests instead to equate energy to
frequency at a fundamental level. In the words of Zeh~\cite{Zeh2009}:
\bquo
In this formulation of quantum theory by means of wave functions, Planck's
constant is {\it not} used primarily to define discrete quantities ("quanta"),
but rather as a scaling parameter, required to replace canonical momenta and
energies by wave lengths and frequencies, respectively, -- just as time is
replaced by length by means of the velocity of light in the theory of
relativity.
\equo
In any case the above Ansatz directly leads to Schr\"odinger's
equation which is known to give the correct physical results,
while the idea of a point particle leads literally to nothing but
problems of both, mathematical and meta-physical nature~\cite{Lyle,Rangacharyulu1997}.

What would be the ``classical'' mathematical alternative to 
Eq.~\ref{eq_correspondence}? What would we have to presume about
the partial waves in order to end up in a distribution of constant 
shape (DCS), which might be closer to a ``classical vision'' of
a particle as some kind of rigid tiny billard ball, as suggested by
Newton in his treatise on optics~\cite{NewtonOpticks}? 
What kind of (linear) wave equation is able to produce such
a DCS? It is well known that any function of the form 
$f(x-v_{ph}\,t)$ solves the homogeneuous wave equation 
$(\d_t^2-v_{ph}^2\,\d_x^2)\,f=0$ 
and produces a DCS, provided that $v_{ph}=\rm{const}$. But this kind 
of wave equation with constant (phase-) velocity represents a
physical situation in which the velocity is a property of the
transmitting medium and not of the moving ``particle''. There was no 
word of such a medium. The desired dispersion does not derive from
the physical properties of some medium: the whole Ansatz derives from
little more than math facts. Therefore Schr\"odinger's theory can be
regarded as a {\it general and abstract} theory of motion for
continuous distributions. Apparently it does not represent the motion
of a specific kind of matter, but of any {\it fundamental} kind of
matter. Here ``fundamental'' means little more than ``not specific''.
We presumed that the matter-density ``represents'' a ``particle'',
i.e. an object which appears to an observer to be fully describable
by it's mass, position and momentum.

It is part of the conventional lore to refuse or even ridicule the
very idea that $\hbar$ could be rationalized (to avoid the
term ``derived'') on classical grounds.
The curriculum taught us to accept $\hbar$ as a joke of nature,
unexplainable by classical rational reasoning. This narrative suggests
that the classical worldview, though it appears to be entirely rational
and logical, is 'de facto' wrong and that quantum physics, though
'de facto' correct, is and will always remain deeply irrational.
But is this fatalistic conclusion, which must eventually shatter the
foundations of enlightenment, namely {\it reason}, and the scientific
enterprize itself, really unavoidable?

There is an oddity related to the often repeated claim
that classical mechanics would necessarily presume point particles: the
mathematical point-mass is among the most discontinuous ideas to be found
in physics, hence in maximal contradiction with a central claim of classical
thought, namely that nature does not jump ({\it natura non facit saltus}).
Therefore a fully self-consistent {\it classical} ontology would have to 
abandon the gospel of the point particle and to consider a smooth 
(differentiable) distribution of matter. This accepted, the above 
reasoning shows that one can rationalize the action constant by applying 
the idea of continuity to the notion of a classical ``particle'': It is
little more than a mathematical expression of a smoothly 
localized substance moving with variable velocity. 

Intuitively it seems evident that some law connecting frequency and wave length
must indeed exist in nature since otherwise traveling waves, i.e. local and
causal signal transport could not exist. But we do not have to refer to
intuition alone, since Toll provided a rigorous general proof of the logical
equivalence of causality and dispersion\cite{Toll}:
\begin{quotation}
[...] the logical equivalence of strict causality and a 
dispersion relation can be expected in any problem in 
which an "output" function is related to a freely 
variable "input" by a linear, bounded, time-invariant
connection. From the invariance of the connection under 
time displacement, it follows" that each frequency 
component is mapped onto itself with only a change in 
magnitude and phase.
\end{quotation}
Hence, to the degree that particle motion is a causal process in time, 
there must be a dispersion relation connected to it.
Consequently the so-called wave-particle ``duality'' is not (only) a
mysterious and alienating property of nature but first of all it is an
unavoidable consequence of math facts about causality and continuity. 
There is no need to regard it as forced upon the theory exclusively by
experimental results: It is a math fact that such a representation must
exist.

It might be surprizing and the consequences difficult to understand, but
it can be derived from classical assumptions and mathematical logic.
Therefore the choice to use (an ensemble of) waves to represent matter
density is not quite as weird as usually claimed: not only
does it allow to cast spatial causality into mathematical form, it might
even be the only way to do so under the presumed conditions.
The dispersion relation $\w(\vec k)$ is, simply by it's 
mathematical form, a Hamiltonian function~\cite{Thorne2017}:
\bquo
The key to geometric optics is the dispersion relation, which 
acts as a hamiltonian for the propagation.
\equo
Hence the ``ensemble'' of partial waves corresponds {\it by mathematical form} 
to an ensemble of ``points'' in an abstract Hamiltonian phase space. 
Hence here we have the connection of two Hamiltonian descriptions: 
The Hamiltonian of a classical particle, the velocity of which must agree 
with another Hamiltonian process, namely the Hamiltonian motion of the real 
and imaginary wave-function components~\cite{Strocchi}.

Mara Beller criticized the rhetoric of ``inevitability'' of the
founding fathers\cite{Beller}. And she had a point insofar as the founding 
fathers claimed inevitability but did not elaborate on it. Hamiltonian notions
provide some evidence that Schr\"odinger's equation has indeed an kind of 
mathematical inevitability. It results from spelling out some simple basic 
principles mathematically, namely continuity, causality and {\it extension}.
Energy quantization is then not a principle at the foundation, it
emerges from the solutions of Schr\"odinger's equation.

This analysis reveals that the invention of an action constant neither 
stems necessarily from discontinuity nor does it contradict causality.
In the contrary, we used both arguments to demonstrate the ``deeper''
logic in Schr\"odinger's equation that presents itself in our view
as a theory of motion for smooth ``ensembles'' (i.e. distributions). 
There is no ``chunking'' of energy but simply a unit conversion factor 
indicating the equivalence of frequency and energy on this supposedly
fundamental level. As Hermann Weyl pointed out, it has the same significance
as the speed of light which implies that it can, using a natural system of
units, it is superfluous. Also John Ralston emphasized that quantum
mechanics can be formulated in a way such that $\hbar$ is
absent and that this approach has many advantages~\cite{Ralstonhbar}.

The true origin of discreteness is not $\hbar$, but was hidden in the
initial assumption:
``Assumed we would, for whatever reason, refuse the notion of  
the point particle and replace it with a (classical) density distribution 
$\rho(t,\vec x)$ in space and time.''
We not only presumed a distributed density, but we presumed that this density
should represent {\it one} particle, i.e. we presumed a
{\it normalization to a discrete number, i.e. to one}. From the perspective
of classical continuum mechanics, the normalization postulate appears as
``non-classical'': Why should the some density represent exactly
``one particle'' and not some arbitrary amount of matter or energy?
On the other hand one can not avoid to admit that Newton did exactly the 
same when he presumed finite discrete particles in his ``Opticks''~\cite{NewtonOpticks}
\bquo
    [...], it seems probable to me, that God in the
    Beginning form'd Matter in solid, massy, hard,
    impenetrable, moveable Particles, of such Sizes
    and Figures, and with such other Properties, and
    in such Proportion to Space, as most conduced to
    the End for which he form'd them; and that these
    primitive Particles being Solids, are incomparably 
    harder than any porous Bodies compounded of them;
    even so very hard, as never to wear or break in pieces;
    no ordinary Power being able to divide what God
    himself made one in the first Creation.
\equo
Hence the postulate of discrete particles is genuinely classical.
It goes back, at least, to the atomistic school of the ancient greek
philosophers. Hence the invention of ``quanta'' can be regarded as
a mathematically maturized version of classical atomistic theory.
The discreteness is not a consequence of $\hbar$. It is more the
other way around: The necessity for $\hbar$ is a consequence of the
discreteness, which was presumed by identifying {\it one} group velocity
with the Hamiltonian equation of motion of {\it one} particle\footnote{
  We smuggled another, inherently Hamiltonian method, into the derivation,
namely the use of complex numbers. We shall come back to this.}.

Precise knowledge of the form of the Hamiltonian function $\w(\vec k)$ is, 
at this point, not required but will be (re-) constructed in the third example.
Note that our second example is similar to the first example in an important
aspect: In the setup example, only half of Hamilton's equation of motion was
given and the mere presumption of the validity of Hamiltonian notions enables 
to derive results that can otherwise only be obtained from a more general
(and apparently distinct) theory, namely Maxwell's electrodynamics.
Again one obtains a physically correct result while it remains 
somewhat unclear why this is so.

From Eq.~\ref{eq_mech_cano} one obtains, for the field free case 
($\phi=0=\vec A$), the de Broglie relations:
\myarray{
{\cal E}&=\hbar\,\w\\
\vec p&=\hbar\,\vec k\,.
\label{eq_deBroglie}
}
This allows to write
\begeq
\psi(t,\vec x)\propto\,\int\,\tilde\psi({\cal E},\vec p)\,\exp{[-i\,({\cal E}\,t-\vec p\cdot\vec x)/\hbar]}\,d^3p\,d{\cal E}\,,
\endeq
which then leads to the ``canonical'' quantization relations
\myarray{
{\cal E}&\to i\,\hbar\,\d_t\\
\vec p&=-i\,\hbar\,\vec\nabla\,.
\label{eq_cquant}
}
If we apply Newton's energy-momentum-relation (EMR) ${\cal E}=\vec p^2/(2\,m)$
for a free particle, then Schr\"odinger's equation of a free particle pops out:
\begeq
i\,\hbar{\d\over\d t}\psi(t,\vec x)=-{\hbar^2\over 2\,m}\,\vec\nabla^2\,\psi(t,\vec x)\,.
\endeq
Combining this with the classical potential energy (density) $\rho(t,\vec
x)\,V(\vec x)$ yields Schr\"odinger's equation of a particle in some 
external potential $V(x)$:
\begeq
i\,\hbar{\d\over\d t}\psi(t,\vec x)=\left(-{\hbar^2\over 2\,m}\,\vec\nabla^2+V(\vec x)\right)\,\psi(t,\vec x)\,.
\label{eq_Schroedinger}
\endeq
Again, though the Hamiltonian formalism involves a causality
requirement, it is difficult to understand why Eq.~\ref{eq_correspondence} 
is in fact more than a ``mathematical trick''. Nonetheless only a single
physical (Hamiltonian) constraint, Eq.~\ref{eq_correspondence}, was
required to arrive at Schr\"odinger's equation. All other steps follow
(auto-/mathe-) matically\footnote{In the established nomenclature of quantum physics, the ``operator'' on the 
right of Eq.~\ref{eq_Schroedinger} is called ``Hamiltonian'' and often the word
{\it operator} is dropped. Since the nomenclature of QM dominates contemporary
physics, we emphasize that we use ``Hamiltonian'' here in the classical sense: 
We refer either to Hamiltonian {\it functions} or, in Sec.~\ref{sec_ex3}, to 
Hamiltonian {\it matrices}. The concept of Hamiltonian {\it operators} emerges
from the above formalism, but is not necessarily fundamental. It is required
to {\it apply} quantum mechanics (QM), but not for it's derivation. }.

\subsection{2nd Example, Aftermath}
\label{sec_ex2x}

The second example contains little that can not be found in standard
textbooks on QM. We only changed the narrative, the presentation of the 
math. This change did not even concern the interpretation of essential
difficulties like the ``measurement problem'', but only our attitude towards
the mathematics of the classical theory.
The major difference has been implemented by the initial sentence 
``Assumed we would, for whatever reasons, reject the notion of a classical 
point particle and replace it with a (classical) density distribution [...]''.

Some textbooks on QM, for instance Messiah's~\cite{Messiah} as well as 
Weinberg's~\cite{Weinberg} and Schiff's~\cite{Schiff} refer to the group velocity
(Eq.~\ref{eq_dispersion}) and the corresponding Hamiltonian expression 
(Eq.~\ref{eq_someHamiltonian}). But all omit to directly derive
Schr\"odinger's equation this way. It is interesting to see what they do instead. 
Messiah first introduces both velocities and writes (page 52):
``From the condition $v=v_g$ and from relation (I.2) one obtains the de Broglie
relations.''
On page 55 he provides another analogy to classical mechanics based on
Fermat's principle. But then, on page 61, one reads the following sentence 
about the possibility to derive Schr\"odinger's equation:
\bquo
It is quite clear that no deductive reasoning can lead us to that equation.
Like all equations of mathematical physics it must be postulated and its
only justification lies in the comparison of its predictions with experimental
results.
\equo
On the same page, Messiah continues to provide three more conditions
that the desired equation must obey, namely a) linearity and homogeneity, b)
first order in time and c) agreement with classical mechanics. On the same
page, he then writes:
\bquo
All these conditions lead us to the Schr\"odinger equation in a natural way.
\equo

With all due respect\footnote{It is not my intention to criticize
Messiah (or Weinberg) specifically. Many ``modern'' textbooks on QM
are {\it de facto} commited to the view that one can not (and/or should not)
make quantum theory plausible at all.}, but these passages send an
inconsistent message: on the one side, we are lead ``in a natural way'' to
Schr\"odinger's equation but, on the other side, it can only be postulated,
for reasons that are ``quite clear''. Whatever Messiah considered to be quite clear,
apparently it was not so clear to those authors who presented derivations of 
Schr\"odinger's equation\footnote{Some examples for such derivations
are to be found in Refs.~\cite{SD1,SD2,SD3a,SD3b,SD4,SD5,SD6a,SD6b,SD7,SD21,SD8,SD9,SD10,
SD11,SD12,SD13,SD14,SD15,SD16,SD17,SD18,SD19,SD20,SD21,SD22,SD23,BRSR2012}. 
This list is long but not exhaustive.} and it apparently was not clear to
P.A.M. Dirac either. Dirac even thought it would have been possible that
Hamilton himself could have possibly stumbled upon quantum
mechanics~\cite{Dirac1963}:
\bquo
Now Schr\"odinger's theory is connected with Hamilton's second development,
the introduction of the families. Schr\"odinger's wave function is
related to Hamilton's principal function $S$. In first approximation
This result has great physical significance. It means that an atomic state
corresponds, not to an individual solution of the classical equations of motion,
but to a family of solutions. The families, which were a mathematical
curiosity at the time of Hamilton, are now seen in their true importance.
Schr\"odinger's form of quantum mechanics may be looked upon as a wave
mechanics which is a natural generalisation of Hamilton's theory of families,
in the same way in which wave optics is a generalisation of geometrical optics.
If Hamilton had known there was any need to generalise the mechanics of
his time, he might have made this step, just by following the optical analogy,
and so have discovered quantum mechanics.
\equo
Also Budiyono and D. Rohrlich seem to promote a position quite
different from that of Messiah~\cite{SD18}:
\begin{quotation}
We cannot fully explain how the theories [quantum and classical mechanics] 
differ until we can derive them within a single axiomatic framework, 
allowing an unambiguous account of how one theory is the limit of the other.
\end{quotation}

Also Weinberg gives a ``historical'' introduction and mentions both, group
velocity and the equivalence with Hamiltonian mechanics on page 14. 
But also Weinberg does not use these relations to derive Schr\"odinger's 
equation.  Though, according to Weinberg (page 21), ``Schr\"odinger showed 
how the principles of matrix mechanics can be derived from those of wave 
mechanics.'' he favors a different approach and writes (page 23):
``The approach that will be adopted when we come to the general principles 
of quantum mechanics in Chapter 3 will be neither matrix mechanics nor
wave mechanics, but a more abstract formulation, that Dirac called 
transformation theory, from which matrix mechanics and wave mechanics 
can both be derived.''
Again students are left with the impression that Schr\"odinger's equation
is somehow important but also somehow impotent. 

Schiff's book also refers to Eq.~\ref{eq_correspondence} (Ref.~\cite{Schiff},
page 17), but speaks of the ``plausibility'' that the de Broglie's relations
receive by it and that there is ``agreement'' found between the group
velocity and classical mechanics due to Eq.~\ref{eq_correspondence}.
Again there is no hint that one could reverse the argument and derive
Planck's constant, de Broglie's famous relations and Schr\"odinger's
equation altogether from the ``assumption'' that physical ``particles''
can't {\it really} be point-like.

Though physicists are usually solicitous to present their science as a 
deductive enterprise, the list of authors that are (supposedly) able but
(apparently) unwilling to derive Schr\"odinger's equation does not stop
here. Cohen-Tannoudji, Diu and Lalo\"e frankly admitted their lack 
of interest~\cite{CohenTannoudji}: 
\bquo
It is possible to introduce it in a very natural way, using the 
Planck and de Broglie relations. Nevertheless, we have no intention of 
proving this fundamental equation, which is called 
Schr\"odinger Equation. We shall simply assume it.
\equo
We have no idea why the authors of a textbook on quantum mechanics could
have {\it no intention} to derive Schr\"odinger's equation other than that
they lost confidence in the value of deductive reasoning. 

Then Messiah surprises (page 6) with the assertion that
``the desire to unify the various branches of their science has always
been one of the most fruitful preoccupations of the physicists'', but
neither his nor other textbooks on quantum theory provide evidence that this 
``preoccupation'' is more than lip-service when it comes to the question
whether one could unify classical and quantum notions. In the contrary, many
textbooks express the pre-occupation that these notions can by no means 
be unified\footnote{Sean Carroll even suspected that physicists actually 
  might {\it not want} to understand quantum theory~\cite{Carroll2019}.}.

There are reasons to doubt that ``point particles'' have {\it ever}
been an uncontroversial ontological element of classical thought. 
In the contrary, a classical lore would rather describe material objects 
as {\it res extensa}, i.e. as objects with extension.
Of course, it is true that textbooks on classical analytical mechanics
{\it describe} particles {\it as if} they had no other properties but a 
``definite'' position, momentum, mass and possibly charge. This
approximation was extremely successful and we agree with F. Rohrlich
who wrote\cite{Rohrlich1996}:
\begin{quotation}
[...] the greatness of these people lay exactly in that fact: that they were 
not deterred by objections no matter how serious they seem to be. Lesser 
scientists may not have dared to proceed in this way. 
\end{quotation}
But this does not imply that there was agreement among physicists of the
classical era that the point particle {\it method} implies any
{\it ontological} commitment, namely that those physicists would have
subscribed to the idea that a particle {\it is} a mathematical point,
i.e. an object without volume. Pre-quantum physicists merely did, what
Rohrlich praised so emphatically: They proceeded despite all known difficulties.

One of the classical natural philosophers who supported (if not introduced)
the gospel of the material point was Boscovich, a Serbian Jesuit 
scientist. Boscovich explicitly postulated material objects without 
extension, likewise called ``material points'' or ``mass points''~\cite{Boscovich}. 
But these ideas were far from being accepted mainstream in classical physics. 
As Glazebrook reported, James Clerk Maxwell was one of the prominent figures 
who rejected Boscovich's view~\cite{Glazebrook}:
\bquo
We make no assumption with respect to the 
nature of the small parts -- whether they are all of one magnitude. We do 
not even assume them to have extension and figure. [...]
The simplest theory we could formulate would be that the molecules behaved like 
elastic spheres, and that the action between any two was a collision following
the laws which we know apply to the collision of elastic
bodies.
\equo
But Maxwell was by far not the only one who refused any ontological claim
about the ``nature of the small parts''.

How about the curriculum? Let us take a look at some arbitrarily
selected\footnote{Most can be found on the internet archive: {\it https://archive.org/}.} 
classical textbooks on analytical mechanics.
In Peck's {\it Elementary Treatise on Analytical Mechanics} from
1887 we find on page 9~\cite{Peck}:
\begin{quotation}
A body is a collection of material 
particles. A body whose dimensions are exceedingly small is called
a material point. In what follows the term point will generally 
be used in this sense.
\end{quotation}
Note that Peck wrote {\it exceedingly small}, but not {\it point-like}.
Twisden explained in 1874~\cite{Twisden}:
\begin{quotation}
A limited portion of matter is called a body e.g. a 
lump of lead is a body. When a body is so small that for 
the purposes of any discussion the relative positions of its 
parts need not be considered, it is spoken of as a material 
pointy or a heavy point, or simply as a point.
\end{quotation}
In Love's textbook on Dynamics, from 1906, we can read~\cite{Love}:
\begin{quotation}
We have said that our object is 
the description of the motions of bodies. The necessity for a 
simplification arises from the fact that, in general, all parts of a 
body have not the same motion, and the simplification we make 
is to consider the motion of so small a portion of a body that 
the differences between the motions of its parts are unimportant. 
How small the portion must be in order that this may be the 
case we cannot say beforehand, but we avoid the difficulty thus 
arising by regarding it as a geometrical point. We think then in 
the first place of the motion of a point.
\end{quotation}
Bartlett wrote 1873~\cite{Bartlett} 
\begin{quotation}
A Primary Property is that without 
which the existence of the body cannot be conceived. There are two of these 
- Extension and Impenetrability.
\end{quotation}
In Michie's textbook from 1887 we read on the first page~\cite{Michie}:
\begin{quotation}
Of the ultimate nature of matter we are ignorant; but from close 
observation of natural laws it has been assumed:
(1) That every material substance is composed of one or more {\it simple
substances} or {\it elements}, so called because they have thus far resisted
simplification by subdivision.
(2) That each of these simple substances is composed of very minute,
but finite and definite, portions, called {\it atoms}.
(3) That in any substance, simple or compound, two or more of these atoms are,
in general, so united as to form the smallest portion that can exist by itself
and remain the same substance. This combination of atoms is called a 
{\it molecule}.
\end{quotation}
No mention of point particles. 
D\"uhring wrote 1873 (page 8)\footnote{ 
``Vor Allem ist es der Begriff des Schwerpunkts, der den ersten 
Ausgangspunkt für die rein theoretischen Untersuchungen
bildet.''~\cite{Duehring}.}:
\begin{quotation}
First of all it is the concept of the {\it centre of
  gravity} that forms the starting point for the theoretical investigation
\end{quotation}
D\"uhring refers to a point in space, namely the {\it center of gravity}.
But of course it would be meaningless to speak of the center of gravity of
point particles: A point {\it has} no center, a point {\it is} a center.
The notion of the {\it center of gravity} makes only sense, if an object
has spatial extension: Analytical mechanics refers to the center of gravity
because this is a successful {\it analytical method}, not because classical
physicists is ``intrinsically'' committed to some point-particle-ontology. 

Wilhelm Schell wrote explicitly (1870) that\footnote{
``Im Uebrigen muss bemerkt werden, dass die Gebilde der Mechanik, 
wie sie hier aufgefasst werden, nur gedachte Dinge sind, wie die 
geometrischen und dass bei der Anwendung der Mechanik auf Vorg\"ange 
der physischen Welt in jedem einzelnen Fall sorgfaltig zu pr\"ufen 
ist, mit welcher Berechtigung und mit welchem Grade der Ann\"aherung 
an die Wirklichkeit man einen physischen K\"orper als ein materielles 
System der einen- oder andern Art ansehen darf.''~\cite{Schell}}:
\begin{quotation}
[...] it must be noted that the objects of mechanics, as
they are described here, are creations of {\it thought} like
geometric figures and that we have to carefully consider
in each single case of application of mechanics to situation
of the physical world to which degree it is legitimate and 
accurate to approximate the reality of physical bodies by 
a system of one type or another.
\end{quotation}
M. Abraham suggested in an article with the title ``basic assumptions of
electron theory'' in 1904\cite{Abraham1904} to understand an electron
as a rigid body of finite size\footnote{
Original in German: ``F. Das Elektron is einer Form\"anderung \"uberhaupt nicht f\"ahig.
G. Es ist eine Kugel mit gleichf\"ormiger Volum- oder Fl\"achenladung.''\cite{Abraham1904}.}: 
\begin{quotation}
F. The electon is not able to change form.\\
G. It is a sphere with homogenuous volume or surface charge.
\end{quotation}
Again there is no mention of an electron as being a ``material point''.

This list is not exhaustive and there might be counter--examples
of physicists who were willing to follow Boscovich. Indeed many textbooks 
on classical mechanics, notably the most ``modern'' ones, do not define
nor discuss the notion of the material point or point particle {\it at all},
but omit any discussion about the value and limits of the mathematical point
as a representation of ``material bodies''.
If they mention these issue at all, then it is usually asserted that
``below a certain radius'', somehow,  classical reasoning has to end and
quantum theory takes over. However, as we have shown, it is mostly the
point-particle-approximation that reached it's limit of validity.

Physicists of the ninetheenth century however, right before (and also
after) Planck's ``quantum-hypothesis'', apparently were aware of the
fact that classical mechanics offered no understanding 
or explanation of the nature of the presumed ``smallest parts''; even 
Boltzmann's kinetic theory of gases was criticized because it required
atoms to exist. The nature of ``particles'' was simply an unexplored area. 
But if the known concepts of mechanics did 
not offer a solution to the particle problem, why should it then be 
a big surprise if it turns out that the old concepts were indeed 
incomplete and insufficient? They were known to be incomplete and
insufficient at all times.

Hence there never really was a general ontological commitment
of classical analytical mechanics to point particles.
Physicists mostly regarded the use of point particle merely as a method
to simplify calculus.
But many textbooks on QM tell this story differently. The general lamento
about the ``difficulties'' with QM suggests that those difficulties
were absent from classical physics, as if classical physics had
been in the posession of some satisfactory theory of matter that 
unexpectedly failed to make the correct predictions. But, as Ralston 
sub-titled, ``there is no classical theory of matter''~\cite{Ralston}. 
Hence the alleged contradiction between classical and quantum 
mechanics misses the point: it is mostly meaningless.
Taken serious, it would mean that an existing and experimetally
confirmed theory, namely Schr\"odinger's wave mechanics, is in conflict
with a theory of material points that actucally never existed.

Some texts, specifically those adressing a non-expert readership,
use the narrative of the classical point particle to underline the
revolutionary new and unexpected content of quantum mechanics --
claiming (for instance) that quantum theory requires that point particles
are ``smeared out'' in space~\cite{Crossroads}. But a particle needs 
to be smeared out in the theory only if it was imagined to be point-like 
in the beginning. Another example of this strange narrative has been given
by Sean Carroll, as he wrote~\cite{Carroll}:
\bquo
    [...] if we think an electron wave function is a diffuse cloud centered
    on the nucleus, when we actually look at it we don't see such a cloud,
    we see a point-like particle at some particular location.
\equo
This is a common assertion. However, it is odd that theoretical
physicists claim to ``see'' other things than experimental
physicists. While Carroll (and many others) claim to ``see'' point-like
particles (which is somewhat difficult to understand), experimentalists 
claim to ``see'' orbitals~\cite{Humphreys1999,Zuo1999,Stodolna2013,Villeneuve2017}.
Of course, since this is in conflict with the (certain readings of the)
Copenhagen doctrine, the latter received harsh criticism since they claimed
to see something that does, according to the theory, not
exist~\cite{Scerri2000}. Even though ``orbitals'' are entirely a
quantum mechanical concept, Scerri wrote~\cite{Scerri2001}:
\bquo
    [...] if these claims [of observed orbitals] were to be sustained it
    would imply an outright refutation of quantum mechanics.
\equo
The experimentalists recanted and admitted that the correct wording
would be ``electron density''~\cite{Zuo2001}. However, the discussion
wasn't finished. Ten years later Labarca and Lombardi discussed once
again ``why orbitals do not exist''~\cite{Labarca2010} and suggested
a split between quantum mechanics and molecular chemistry:
\bquo
Therefore, the quantum world has no priority over the world of molecular
chemistry: chemical entities do not need the support of quantum
entities to legitimate their objective existence. From this perspective,
orbitals exist in the ontology of molecular chemistry, in spite of the
fact that they do not exist in the quantum world.
\equo
Hence, if quantum theorists start to work in molecular chemistry (or vice
versa), apparently they should receive a training in doublethink beforehand.

But not only classical analytical mechanics avoided a commitment to
point particles. Sebens wrote recently~\cite{Sebens}:
\bquo
[..] we examined some of the reasons why it is appealing to think of electron
charge as spread out in the way Schr\"odinger proposed. [...]
Although quantum chemists regularly treat wave functions as describing spread-out
distributions of charge, scholars working on the foundations of quantum mechanics
rarely explicitly include such charge densities in the ontologically precise formulations
of quantum mechanics that they propose. [...] Here I have argued that their
fit with quantum chemistry is a point in their favor. When we move to quantum
field theory, I think the case for a spread-out electron charge density is
particularly strong as the theory can be viewed as describing
quantum superpositions of classical field states where electron charge is
spread-out.
\equo
And Rangacharyulu, 1997~\cite{Rangacharyulu1997}:
\bquo
[...] in microscopic physics the discussions of wave-particle
duality are not meaningful. This reasoning is based on the observation that
waves are a conglomeration of coherent disturbances in a many body system and
as such they do not represent individual entities. Newtonian kinematics have
no predictive power and they do not offer a physical description of
participants in an interaction, except to say that they obey the conservation
laws. The point-particle concept is an unnecessary complication in physics.
\equo
Or S.N. Lyle~\cite{Lyle}:
\bquo
The point particle approximation has been extraordinarily successful.
But [...] we might understand physics better by knowing what can be done with
spatially extended particles.
\equo
Even the prominent string theorist M. Kaku regards it as an achievement
that strings are not point particles~\cite{Kaku}:
\bquo
From a technical point of view, superstring theory seems to be totally free of
quantum anomalies and divergences, which riddle all known point-particle
theories of gravity and matter. [...]
In quantum field theory, point-particles interact via Feynman graphs,
which badly diverge when the graph is ``pinched,'' that is, when one of
the legs of the graph shrinks to zero. When the string moves, it repeatedly
splits and reforms, thereby tracing the topology of two-dimensional sheets
or Riemann surfaces, such as a doughnut. However, since it is difficult
to pinch or stretch a doughnut, one can show that the string graphs are
actually ultraviolet finite. Thus, the topology of the string removes the
divergences of quantum gravity.
\equo
This does not imply that we intend to promote string theory. But it should
be noted that the point particle imagery does not enjoy universal support,
neither in classical nor in post-classical physics.

Our small literature survey of classical textbooks reveiled a variety of
suggestions of how particles should be understood, some more pragmatic and
some which are more philosophically minded. But there is no support for
the narrative that the classical worldview of the 
nineteenth century is unequivocally based on ``material points'' 
in an ontological sense. None of the textbooks on analytical 
mechanics that we found contained ontological claims about the
nature of particles. The author does not claim deep expertise in the 
history of science, but it seems to us that the insistence on the alleged
``classical'' point-particle is mostly an invention of the 20th century,
maybe with the (somewhat legitimate) intention to underline the revolutionary 
content of quantum theory.

But if classicality requires particles - but not {\it point}-particles, then
few arguments are left to deny that Schr\"odinger's equation is {\it as such}
perfectly classical: it provides a mathematical description of distributed
normalized matter density moving in space constrained by a classical
Hamiltonian dispersion relation. It provides a continuity equation and hence
obeys a local conservation law. It is, taken as such, in any reasonable
sense of the word a classical theory. Not only does the presumption that
particles can't
be represented by mathematical points allow to ``naturally'' arrive
at Schroedinger's equation, it has also been shown elsewhere that the
classical limit of free wave packets, often suggested to be restored by
$\hbar\to 0$, indeed reproduces classical point mechanics~\cite{qm2cm}:
\begin{quotation}
In the limit $\hbar\to 0$, the extension of the Gaussian wave packet
considered in Refs. 1 and 4 approaches zero in both coordinate space and
momentum space.
\end{quotation}

It is often taught that free wave-packets are always spreading out in
space and that wave-packets are therefore somehow non-classical.
However, it has been shown that wave-packets do not {\it always}
spread out~\cite{wfspreading}, but that their width $\Delta x$ is
actually given by
\myarray{
  \Delta x_t&=\sqrt{\Delta x_0^2+A\,t/m+B\,t^2/m^2}\\
  A&=\langle{\bf x}\,{\bf p}+{\bf p}\,{\bf x}\rangle_0-2\,\langle{\bf x}_0\rangle\,\langle{\bf p}\rangle\\
  B&=\delta p^2\\
\label{eq_uncertainty_qm}
}
This means that they spread out {\it eventually} but, since $A$ can be negative, 
not necessarily at all times: They might be convergent in the beginning.
Let us compare this result to a bunch of classical particles with 
${\bf x}_i=\langle{\bf x}\rangle+{\bf\delta x}_i$
and ${\bf p}_i=\langle{\bf p}\rangle+{\bf\delta p}_i$. In the force-free
case we have (for every direction $x,y,z$):
\myarray{
  x_i(t)&=x_i(0)+p_i/m\,t\\
\langle x\rangle(t)&=\langle x\rangle(0)+\langle p\rangle/m\,t\\
\delta x_i(t)&=x_i(0)+p_i/m\,t-\langle x\rangle(0)-\langle p\rangle/m\,t\\
             &=\delta x_i(0)+\delta p_i/m\,t\\
}
The square is:
\myarray{
\delta x_i^2(t)&=(\delta x_i(0)+\delta p_i/m\,t)^2\\
&=\delta x_i^2(0)+2\,\delta x_i(0)\,\delta p_i/m\,t+\delta p_i^2/m^2\,t^2\\
\langle\delta x^2\rangle(t)&=\langle\delta x^2\rangle(0)+2\,\langle\delta
x\delta p\rangle(0)/m\,t+\langle\delta p^2\rangle/m^2\,t^2\\
\label{eq_uncertainty_cm}
}
with 
\myarray{
  A&=2\,\langle\delta x\delta p\rangle(0)\\
  &=\langle x\,p+p\,x\rangle-2\,\langle x\rangle_0\,\langle p\rangle\\
}
this expression is identical to the quantum-mechanical result.
Hence the wave-packet expands exactly as an ensemble of straight trajectories
(a beam of drifting particles) expands -- no more and no less. 

In any real--world experiment, the precision with which we ``prepare''
a particle with a specific momentum at a certain position, is always finite.
How do we (classically) ``prepare'' particles in a certain state of motion?
Most likely one does not {\it prepare} at all, but {\it select}: 
Typically one would use a beam of particles and select the right particles
by two (or more) small-aperture collimators in order to define angle and position 
within specific ranges. The number of particles able to pass two distant 
collimators, i.e. the intensity of the transmitted beam, then is usually
proportional to the area of the apertures and to the range of spatial angles,
because the incoming beam, stemming from a source of finite temperature, 
will have a thermal statistical distribution.
This product of area and momentum range is called the phase space volume 
and the intensity of a beam of particles is proportional to the phase space
volume that the setup of collimators accepts. Hence infinite precision is
mathematically possible but implies zero particle flow and is hence
{\it physically meaningless}, both in classical mechanics as well as in
quantum theory.

Now quantum mechanics has a bit more to say, namely that particles are 
``res extensa'' in phase space.
This means that ``a particle'' cannot be defined by it's finite 
spatial volume alone: if material particles are {\it res extensa}, this concerns 
phase space: spatial volume is {\it always} just one factor, the other being 
the volume in momentum-space\footnote{It is one of the absurd consequences of 
the Newtonian curriculum that many post-graduate students are able to play back 
the Copenhagen gospel but have not been told about (or not absorbed) the
concept of phase space.}. From a mathematical perspective, the problem to
squeeze quantum particles through tiny holes, is (disregarding interference)
comparable to the problem of squeezing a statistical ensemble of point
particles through tiny holes. It can only be effective with a ``convergent''
wave-function, which is identical to a large negative correlation of position
and momentum (see Eq.~\ref{eq_uncertainty_qm}). The exact same Eq.~\ref{eq_uncertainty_cm} 
holds for a statistical ensemble of particles: The smaller the region where 
to concentrate the beam, the steeper the required focusing\footnote{
Focusing means exactly this: to generate a large negative correlation of
(local) position and momentum.}. 
That's basically a supplement to the content of Liouville's theorem: we can't 
squeeze, even a single particle, into a phase space below a certain volume~\cite{Gosson}.

The (partial) equivalence (and competition) between particle- and wave-picture 
was known for long in classical optics. The issue was thought to be solved by
Young's double-slit experiment in 1801: Light shows interference effects
and hence ``is'' a wave. However, the proof was based, as we know today, on 
an incomplete understanding of matter. Since Young's double slit experiment
has meanwhile been successfully repeated with matter-waves, we have to review
it's original goal and conclusion. It should be clear today that Young's
double slit experiment actually did {\it not} decide the issue: {\it There
  never really was an issue to decide}. The conviction that particles can not
interfere is obviously erroneous, because it was derived from a flawed
concepts of particles and their motion: Schr\"odinger's theory can be regarded
as the first causal theory of the motion of (distributed) matter. 

Based on {\it macroscopic} experience with matter classical physicists had 
wrongly concluded that there is a fundamental difference between 
waves and {\it microscopic} particles. But the physical properties of 
macroscopic solid bodies are emergent; they are (obviously) not equal to
the fundamental properties of individual ``particles''. The projection of the
macroscopic notions down to the dimension of single ``particles'' created
the impression of a dichotomy that only exists in our imagination. This is the
straightforward lesson to be drawn from the success of Schr\"odinger's theory.

Heisenberg remarked that Eq.~\ref{eq_density} could be normalized to any
value and not just to the number of particles~\cite{Heisenberg0}. This is
of course correct but it makes no difference in principle whether one claims
that the density must be normalized to unity because it is one particle that
is described or because it is the probability density ``to find one
particle''. The difference is merely in verbiage, not in
(mathematical) essence. In both cases it is the act of 
normalization that generates the real ``quantization'' of matter by
declaring that some function represents one particle.

Nonetheless it should be emphasized that the ``classical'' perspective described 
here neither originates nor requires specific metaphysical presumptions.
The mentioned requirements of finality, continuity and causality, no matter the
metaphysical interpretation of the density, are sufficient for the
mathematical derivation of Schr\"odinger's equation.

Hence, up to this point we did not (need to) specify what the density $\rho$ 
actually represents. It might still be open for philosophers to argue
for a ``real'' matter density or ``just'' a probability density. However,
in view of the mentioned math it makes few sense to describe the squared
wave function as a probability to find a ``point particle''.

Whatever philosophy will eventually prevail, it should be noted that
Max Born, awarded the Nobel prize for his probabilistic interpretation
of the wave-function, wrote\cite{Born1956}:
\begin{quotation}
though the wave functions are representing, by their square, probabilities,
they have a character of reality. That probability has some kind of reality
cannot be denied. How could, otherwise, a prediction based on probability
calculus have any application to the real world?
\end{quotation}
Hence even Born, a virulent defender of the Copenhagen philosophy, could
not believe in it's most obscure anti-realistic gospel: Whatever the wave
function represents, it can not fully be abstracted from reality if it is
supposed to represent (aspects of) this reality.
Maybe there is, after all, no reason to scandalize the appearance of probability
densities, nor to scandalize the imagery of electron--waves. Probabilites
are known in classical mechanics as well and the ``difficulties'' to
decide between waves and particles are known from classical physics as
well. Quantum theory is not a new disease but the cure of a long-standing
problem that physics was not aware of: How to rationalize motion of extended
matter in space in terms of strict local causality.

Newton's axiom that force free bodies 
move rectilinear presumes that ``motion'' is a self-evident process that 
needs no further explanation~\footnote{His theory requires, however, the
opaque notion of inertia, yet another unsolved riddle of Newtonian mechanics.}.
Apparently Schr\"odinger's equation mostly provides a new and different account
of motion\footnote{See also Ref.~\cite{Cohen2017,Dethe2019}.}. 
However, in light of Toll's proof, the Schr\"odinger equation is indeed
the simplest causal theory of the motion of a distributed portion of
matter in space - using a minimal set of specifications.

For those who study the history of quantum mechanics and in view of this
fairly simple and straightforward logic, the question may arise, what
kind of rationalization the ``founding fathers'' actually
had in mind when they proclaimd that the Copenhagen interpretation
clarified everything.

Weizs\"acker recalled an interesting conversation (see page 184 in Ref.~\cite{Weizsaecker}):
\begin{quotation}
With him [Heisenberg] I existed in a state of tension such as can only
arise when one is very close to another person. In Berlin in April 1927,
in a taxi, he told me of the uncertainty principle saying, "I think I
have refuted the law of causality"; in that moment I decided to study
physics to understand this."
\end{quotation}
This quote refers to the celebrated ``uncertainty paper'' and testifies
Heisenberg's enormous ambitions but also raises doubts concerning the 
strength of his scientific sobriety that might have prevented him
from drawing quickly bold conclusions.
These doubts are also due to what MacKinnon reported~\cite{MacKinnon}:
\begin{quotation}
When Bohr returned from Norway he read the [draft of the uncertainty] paper 
and thought that it should be treated the way initial drafts of his own 
papers were. It should serve as a basis for discussion and be written and 
rewritten until every detail was correct.
Heisenberg ignored all such suggestions and sent the hastily written paper,
with all its imperfections, to the Zeitschrift f\"ur Physik. The indeterminacy
principle decisively undercut Schr\"odinger's wave picture, which in princi-
ple assumed precise specifications of both position and momentum. 
Heisenberg wanted his paper published as soon as possible.
\end{quotation}
Heisenberg himself described his motivation to demolish
Schr\"odinger's competing interpretation quite frankly~\cite{PAB}:
\bquo
Now Schr\"odinger's interpretation -- and this was the great novelty --
simply denied the existience of these discontinuities. [...]
This hypotheses seemed to me too good to be true, and I mustered what
arguments I could to show that discontinuities were a fact of life,
however inconvenient.
\equo
This quote contains a central point that requires critical review:
We are not aware of any physical method which would allow to verify
true discontinuities as a ``fact of life''.
As we shall argue in part two, true discontinuities have the intrinsic
property that they can neither be verified nor falsified by experiment.
It is therefore questionable whether it is a legitimate scientific
hypothesis, after all.

Schroedinger's theory, though derivable from classical logic,
by far exceeds the scope of the classical theory. It generates new 
features as compared to the conventional classical treatment of motion. 
His theory introduces a second levels of superposition: 
The superposition of the ``auxiliary'' functions $\psi$ and 
of the densities $\rho=\psi\psi^\star$. This implies some ``non-classical'' 
features. The (linear) superposition of densities is given by\footnote{
See also Ref.~\cite{BricmontQSNS}.}
\begeq
\rho_1+\rho_2=\psi_1\psi_1^\star+\psi_2\psi_2^\star\,,
\endeq
while the superposition of wave-functions yields
\begeq
(\psi_1+\psi_2)\,(\psi_1+\psi_2)^\star=\rho_1+\rho_2+\psi_1\psi_2^\star+\psi_2\psi_1^\star
\endeq
Both equations can only agree if the wave functions don't overlap 
or if the product of the wavefunction is skew-symmetric with respect 
to an exchange of the ``particle's'' index.
Hence these wave functions cannot be superimposed arbitrarily\footnote{
This feature, when projected backwards to the idea of classical 
``point particles'', remains a mystery. But it is entirely 
comprehensible in the wave picture.} and if one simply scales the 
normalization of the wave-function to represent two particles 
instead of one, one breaks the rules suggested by the dispersion
relation: For two particles, two dispersion relations are required 
in order to obtain two velocities. Hence the ``configuration space'' 
of the partial waves must be of a higher dimension. Furthermore, 
from a classical viewpoint, there is few reason to expect that 
density-distributions are intrinsically additive unless the 
represented particles don't interact {\it at all}. 

A new perspective, even if it is based on classical logic, may have 
unexpected consequences that go beyond the range of ``classical'' reasoning.
If we prefer the view that all consequences of ``classical'' thought 
are by definition classical results, then Schr\"odinger's equation is
classical -- and that was according to all known sources his own initial
understanding. But if we prefer to say that quantum theory begins with Planck's 
constant, then the wave-particle duality of Eq.~\ref{eq_correspondence} 
can be regarded as the {\it logical} origin of quantum theory. Everything
that follows from it is ``quantum'', no matter how we interpret its content. 
Maybe the distinction between ``classical'' and ``quantum'' is rather 
a matter of philosophy than a matter of mathematical logic~\cite{RethinkCP}:
\bquo
What is ‘classical physics’? Physicists have typically treated it as a useful and unproblematic
category to characterize their discipline from Newton until the advent of ‘modern physics’ in the
early twentieth century. But from the historian’s point of view, over the last three decades
several major interpretive difficulties have become apparent, not least the absence of
unequivocal criteria for labelling physicists and their work as ‘classical’, whether during the
nineteenth century or earlier. Some historians have consequently either treated the term as a
retrospectively contrived anachronism (such as Olivier Darrigol), or carefully avoided using it in
their analyses (such as Jed Buchwald). Nevertheless, current historiographies have not
systematically explored the implications of abandoning ‘classical physics’ as an analytical
category. As a result, they arguably overstate the unity of the physics prior to the rise of quantum
and relativity theories in the twentieth century. Moreover, many studies into the activities of late
nineteenth-century physicists have adopted the perspective of later theoretical developments
typically associated with the birth of one type or another of ‘modern physics’, for example the
origins of microphysics and, through special relativity, the history of electrodynamics. This
focus on theoretical discontinuities, implicit in the classical/modern distinction, has long diverted
attention away from important historical continuities in both experimental practice and the
applications of physics. We take these reasons as sufficient motivation for rethinking ‘classical
physics’.
\equo

The main theme of this article is to show the amazing power of 
Hamiltonian notions in physics. Even if we follow them almost
blindly, apparently they can guide us to new valuable insights. 
We think that Schr\"odinger's equation not only provides an excellent 
example for the fruitful use of Hamiltonian methods, but is also an 
appropriate and possibly necessary introduction to the next example.
Influenced by Whittaker's {\it Treatise on the Analytic Dynamics}, it was
Dirac's explicit intention to keep as much as possible of the classical
Hamiltonian notions~\cite{Schweber1994}. As we shall argue in the next
example, Dirac's theory can even be regarded as a Hamiltonization of
space-time geometry.

\section{Third Example: Dirac's Equation}
\label{sec_ex3}

\bquo
The discovery of this [Dirac's] equation was the most important advance
in the theory of the electron since the Maxwell-Lorentz equations of
classical electrodynamics. Bohr's semiclassical theory and non-relativistic
quantum mechanics served only as transitional theories.\\
\mbox{}\hfill{\it -- Sokolov, Loskutow and Ternov}~\cite{SLT1964}
\equo
Dispersion relations are well known to mathematicians, engineers and 
physicists~\cite{Hamilton,Brillouin,Whitham,Pain,Toll,Guettinger,D4GHTY}.
We repeat that Toll proved the ``logical equivalence of strict causality
and a dispersion relation''~\cite{Toll}.
However, from the Hamiltonian perspective, it is disturbing that we
had to use Newton's EMR to arrive at Schr\"odinger's equation. 
We might better argue that the frequency must, for reasons of isotropy, 
be an even function of the wave-vector and therefore 
must have a Taylor series expansion $\w(\vec k)=c_0+c_2\,\vec k^2/2+\dots$.
This is a strong argument, based on a minimum specification, to establish
Newton's dispersion relation. But we must be willing to follow Newton's
method and to presuppose 3-dim. isotropic absolute space. But there is
an alternative to this Newtonian logic. We suggest to follow a purely
Hamiltonian way of thought, since this appears to provide reason for the 
Newtonian approximation and for the need to replace Schr\"odinger's by
Dirac's equation.

So the question remains: where does the relativistic dispersion
relation (RDR) come from?
In the conventional 'historical' account, it is not derived from Dirac's 
equation, but rather the other way around: Dirac somehow 'guessed' his 
equation in order to reproduce the RDR with a first order (i.e. Hamiltonian) 
equation. The conventional lore suggests that we have to presume the 
constancy of the speed of light and the ``metric'' of Minkowski space-time 
in order to arrive at the RDR. But is that the only possibility? 
Do we {\it have to} speak about ``inertial frames''
and ``clock synchronizations'' in the first place\footnote{
Einstein himself was not fully satisfied with the notion of the
{\it inertial frame}. In a letter to Jaffe he wrote in 1954:
``I see the most essential thing in the overcoming of the inertial 
system, a thing which acts upon all processes, but undergoes no 
reaction. The concept is in principle no better than that of the 
centre of the universe in Aristotelian physics''~\cite{Stachel}.}?

The common presentation of special relativity rests on two central
pillars, the constancy of the speed of light and the invariance
of the laws of physics. The latter principle, however, namely the
invariance of Hamilton's equations of motion under canonical
transformations, is already integral part of Hamiltonian notions
and the theory of canonical transformations.
It might be defendable as a fundamental requirement for a theory
that is point-of-view-invariant~\cite{StengerBook}.

But the postulate of the constancy of the speed of light suggests
that some {\it specific} physical wave phenomenon, namely light, 
serves as a {\it general} limit that determines the relative scale 
of spatial vs. temporal coordinates. This implies that this 
{\it specific} physical phenomenon has {\it fundamental} physical 
significance, i.e. that it is in fact not a {\it specific} phenomenon 
but instead a {\it general} principle, that does not {\it logically} 
originate in the transmission of light.
But, to our knowledge, the theory of special relativity does not explain 
{\it why} electromagnetic phenomena should play this fundamental role 
since the standard approach regards electromagnetism as one of four 
fundamental ``forces of nature'', i.e. as one among several others. 
There is no explanation given why the propagation of light should have
any {\it special} significance in a space-time theory. It is therefore
not surprizing that many scientists seeked for alternative foundations
of the Lorentz transformations and relativity~\cite{Ignatowski1,Ignatowski2,Drake,Eisenberg,
Alstroem,BacryLL,Berzi,Lee,LevyLeblond,Mermin84,Schwartz,
Field,Huang,Pelissetto,MMC,Chao1,Chao2,Gao,DaiDai,Mathews,Duarte}. 

The question then is: is it required to raise the mere phenomelogical 
fact of the constancy of the speed of light to the status of a 
fundamental principle? As it is mere an empirical fact about nature, 
it should be an outcome of the theory, and not be placed at the very foundation. 
We shall show in what follows that the relativistic dispersion relation 
can be indeed be derived from Hamiltonian symmetries.
This requires to use some bits of linear algebra, but though many 
math facts also hold in more general cases, we have to apply them 
to nothing more demanding than real $4\times 4$-matrices. The sequence
of arguments that allows to derive the Dirac algebra and the relativistic
dispersion relation from pure Hamiltonian arguments is long but 
quite rigorous~\cite{qed_paper,osc_paper}.

Dirac's idea to implement the RDR by matrices, as ingenious as it was,
was yet another ad-hoc approach that pre-supposes the Minkowski metric and 
leaves physics students with new riddles: what is the meaning of spinors and 
the Dirac matrices? Why do we need to multiply the Dirac spinor by $\y_0$
to obtain the ``adjunct'' spinor? What is the sense and significance
of $\y_0$ anyway? What are the arguments that might lead to a 
``space-time metric'' which is not positive (semi-) definite?
 
David Hestenes was probably the first who recognized the elegance of an
algebraic description of geometry on the basis of the Dirac-Clifford algebra.
He reformulated Dirac's theory and developed the {\it space-time algebra} (STA), 
as he called it. This algebra reveals the deeper connection between Dirac's
theory, the geometry of space-time and electrodynamics~\cite{STA}. We think
that this was an essential step forward without which it is impossible to
grasp the dynamical emergence of space-time.
But to our knowledge Hestenes made no attempt within his theory to explain 
why this specific algebra (and not some other) should be ideally 
suited to describe space-time and electrodynamics: why is the Dirac algebra 
fundamental?

Hence there is still a missing bit, a missing logical reason for why things 
are as they are and not some other way. There are good arguments supporting
the view that this missing bit is the connection of the Dirac equation
with ``classical'' symplectic motion\footnote{We use quotes in
  ``classical'', because the symplectic group was
  introduced by Hermann Weyl~\cite{Weyl1939} in 1939 and 
was therefore unknown when Dirac formulated his relativistic theory of 
the electron. Though the symplectic group originally seemed to belong to 
the realm of classical physics, the intrinsic connection to quantum mechanics 
is known and has been elatorated in some detail; see
Ref.~\cite{Jost1965,Ralston1989,BB,Gosson}.}.
Though the insight that quantum theory comes along with a symplectic structure 
is not new, it has long been overlooked that the Dirac algebra itself can
be derived from purely Hamiltonian notions, not only by analogy, but
literally~\cite{qed_paper,osc_paper,lt_paper}. Even this is no new insight
but had been understood, for instance by Kim and Noz, already fourty years
ago~\cite{Kim81}:
\begin{quotation}
  From a mathematical standpoint, special relativity is the physics
  of Lorentz transformation, and quantum mechanics is the physics of
  Fourier transformation. It is easy to see, if not well known,
  that the Lorentz boost is a symplectic transformation in the plane
  of longitudinal and timelike coordinates. In Fourier transformation,
  the width of the momentum distribution is inversely proportional to
  that of the spatial distribution. This is also a transformation
  property of the symplectic group. Thus, it is not unreasonable
  to suspect that the natural language of relativistic quantum mechanics
  is the symplectic group.
\end{quotation}

Dirac's algebra is not only useful to describe the geometry of Minkowski's
space-time and relativistic quantum mechanics. The real Dirac algebra
provides a general parameterization of arbitrary real $4\times 4$-matrices.
This parameterization, however, not only matches the requirements of
Lorentz covariance, it furthermore allows to derive the Minkowski metric:
the real Dirac algebra provides an optimal parameterization 
of the Lie algebra of the symplectic group $sp(2n=4,\mathbb{R})$ which 
generates the full set of possible linear transformations between two 
canonical pairs~\cite{rdm_paper,geo_paper,stat_paper,jacobi_paper}.
In other words: The Dirac algebra pops out of a formal analysis of 
classical Hamiltonian mechanics and therefore, it is {\it intrinsically}
related to the notion of ensembles in classical Hamiltonian phase space. 
Hence if the partial waves that have been used in Schr\"odinger's equation
are indeed points in a Hamiltonian phase space, then the dispersion
relation should have it's logical origin in Hamiltonian notions.

The wave-function, i.e. the Dirac spinor, corresponds in the momentum 
representation to ensembles of two classical oscillators, 
just as Schr\"odinger's equation suggests and in perfect agreement with
Dirac's understanding of what the components of his spinor actually 
represent~\cite{Dirac}: 
\begin{quotation}
These new degrees of freedom are to be associated here with certain 
dynamical variables $(q_1,p_1)$ and $(q_2,p_2)$ to be thought of as 
corresponding to two independent linear harmonic oscillators. 
\end{quotation}
But what can be said about two such oscillators without a
specific description of the oscillating system, for instance in terms 
of masses and spring constants? A different type of analysis is needed,
a kind of ``contentless deductive theory''~\cite{Loewdin}:
\bquo
If the elements of the group are not given any realization, and the group is
essentially defined by its multiplication table, one obtains an abstract group
theory, which may have realizations ranging from atomic physics to
oriental carpets.
\equo
That is, instead of starting from a metaphysical theory of how space, 
time, matter and fields should be understood, we start with an empty pure 
and general Hamiltonian {\it form}.

One then has to proceed, lacking any phenomenological context and further
specification, with a survey of the full space of {\it possible} solutions
of arbitrary general Hamiltonian functions of an arbitrary number of
canonical pairs: What are the general features of such systems which
might explain how some general kind of dispersion relation emerges? Here
we take the classical path and start with a general Hamiltonian function
in the form of a Taylor series in the dynamical variables. We then apply
some simplifying assumptions.
These assumptions should not be misunderstood as metaphysical presuppositions.
It is simply the approved method in the history of physics to attack a
problem of this dimension and generality by considering the simplest
possible cases first. This might help to develop the notions required to
elaborate solutions for more complex situations. Hence we start with a
second order approximation for a Hamiltonian of $n$ canonical pairs, i.e.
with the most general form of linear equations of motion in $2\,n$ variables.
The simplest non-trivial Hamiltonian system is a canonical pair and the simplest
system with coupling consists of two canonical pairs. It should provide us 
with an understanding of the {\it principle form} of linear Hamiltonian couplings.

Before we enter the details, let us emphasize that it is not as far-fetched
as it might appear at first sight to relate linear Hamiltonian couplings
to wave mechanics. The difference between an ensemble of non-interacting
oscillators and a linear chain 
-- and hence wave motion -- lies in the coupling between oscillators: 
Waves are, in a very general sense, the result of {\it coupled}
oscillations~\cite{Pain}. Hence it is reasonable that the {\it general} 
algebraic structure of the coupling determines the {\it general} 
characteristics of wave motion, namely of the motion that we expect
to generate a physically meaningful purely Hamiltonian dispersion relation. 
The simplest oscillator is represented in Hamiltonian theory by a single 
canonical pair, hence the simplest coupling requires two canonical pairs.

Let $\psi=(q_1,p_1,q_2,p_2)^T$ represent Dirac's two classical canonical 
pairs (two degrees of freedom), then the quadratic terms of a
general Hamiltonian function are given by
\begeq
{\cal H}=\frac{1}{2}\,\psi^T\,{\bf A}\,\psi
\label{eq_linH}
\endeq
where ${\bf A}$ is a positive definite real symmetric $4\times 4$ matrix. 
We restrict our considerations to symmetric matrices since skew-symmetric
components do not contribute to the Hamiltonian function.

The Hamiltonian function is a constant of motion if
\begeq
{d{\cal H}\over d\tau}=\dot{\cal H}=(\nabla_\psi{\cal H})\cdot\dot\psi=(\psi^T\,{\bf
  A})\cdot\dot\psi=0\,,
\label{eq_conservation}
\endeq
which has the general solution
\begeq
\dot\psi=\y_0\,\nabla_\psi{\cal H}=\y_0\,{\bf A}\,\psi={\bf F}\,\psi\,.
\label{eq_eqom}
\endeq
where $\y_0$ is a skew-symmetric matrix, the so-called 
{\it symplectic unit matrix} (SUM). In principle this matrix 
could have an arbitary skew-symmetric form, but it can (with some mild and
reasonable assumptions) be brought into the following form~\cite{MHO}: 
\begeq
\y_0=\bmtx{cccc}
0&1&0&0\\
-1&0&0&0\\
0&0&0&1\\
0&0&-1&0\\
\emtx
\endeq
Hence the SUM $\y_0$ implements Hamilton's equations of motion in algebraic form,
which becomes obvious if one writes Eq.~\ref{eq_eqom} explicitely in components:
\begeq
\bmtx{c}
\dot q_1\\
\dot p_1\\
\dot q_2\\
\dot p_2\emtx=\bmtx{cccc}
0&1&0&0\\
-1&0&0&0\\
0&0&0&1\\
0&0&-1&0\\
\emtx\,\bmtx{c}
{\d{\cal H}\over\d q_1}\\
{\d{\cal H}\over\d p_1}\\
{\d{\cal H}\over\d q_2}\\
{\d{\cal H}\over\d p_2}\\
\emtx
\endeq
Remarkaby, the skew-symmetry of $\y_0$ alone suffices to qualify
Eq.~\ref{eq_eqom} as a solution for Eq.~\ref{eq_conservation}.
 
Matrices of the form ${\bf F}=\y_0\,{\bf A}$ are called Hamiltonian
and they are the starting point of linear Hamiltonian theory. More generally,
a matrix ${\bf F}$ is said to be {\it Hamiltonian}, iff it obeys~\cite{MHO}
\begeq
\y_0\,{\bf F}\,\y_0={\bf F}^T\,.
\label{eq_symplex}
\endeq
It is not immediately obvious from Eq.~\ref{eq_symplex}, but in combination
with $\y_0^2=-{\bf 1}$ and $\y_0=-\y_0^T$, Eq.~\ref{eq_symplex} combines
matrix transposition with commutation relations. Two matrices ${\bf A}$ and ${\bf B}$
are said to commute, if ${\bf A}\,{\bf B}-{\bf B}\,{\bf A}=0$ and to
anti-commute, if ${\bf A}\,{\bf B}+{\bf B}\,{\bf A}=0$. Eq.~\ref{eq_symplex}
allows to construct two matrices ${\bf F}_{a,c}$ such that ${\bf F}_a$ anti-commutes
with $\y_0$ while ${\bf F}_c$ commutes with $\y_0$:
\myarray{
{\bf F}_a&={\bf F}+\y_0\,{\bf F}\,\y_0\\
{\bf F}_c&={\bf F}-\y_0\,{\bf F}\,\y_0\\
\label{eq_ac}
}
The original matrix is ${\bf F}=({\bf F}_a+{\bf F}_c)/2$.
Inserting Eq.~\ref{eq_symplex} into Eq.~\ref{eq_ac} results in
\myarray{
{\bf F}_a&={\bf F}+{\bf F}^T\\
{\bf F}_c&={\bf F}-{\bf F}^T\\
\label{eq_ac_sym}
}
such that ${\bf F}_a$ is symmetric and ${\bf F}_c$ is skew-symmetric.
Hence Hamiltonian matrices that commute with the SUM $\y_0$, are 
skew-symmetric and those that anti-commute with $\y_0$, are symmetric.

The general solution of Eq.~\ref{eq_eqom} for constant ${\bf F}$ is
given by the matrix exponent ${\bf M}(\tau)$ 
\begeq
\psi(\tau)=\exp{({\bf F}\,\tau)}\,\psi(0)={\bf M}(\tau)\,\psi(0)\,.
\label{eq_motion}
\endeq
It is a math fact that ${\bf M}$ is a symplectic matrix, iff ${\bf F}$
is Hamiltonian. The evolution in time, generated by some 
Hamiltonian matrix ${\bf F}$, is a sympletic (canonical) transformation.
One can show that symplectic matrices obey~\cite{MHO}:
\begeq
{\bf M}\,\y_0\,{\bf M}^T=\y_0\,.
\label{eq_symplectic}
\endeq
Symplectic matrices form a group which means that any product of
symplectic matrices is again a symplectic matrix.

In Hamiltonian theory observables are generators of canonical
transformations\footnote{
Andre Heslot\cite{Heslot1983}: 
\begin{quotation}
Current textbooks often emphasize the generator aspect of observables in
quantum mechanics, but it is seldom mentioned that this aspect already exists
in classical mechanics. As a consequence, notions which already make sense in
classical theory are too often considered as purely quantum ones: The spin is
a striking example of such a confusion. As in quantum mechanics, the generator
aspect of observables in classical mechanics may be dealt with by
mathematically sophisticated group theoretical considerations, but we hope we
have convinced the reader that this aspect proves relevant even at the
elementary level.
\end{quotation}
}. 
So what are the observables and how do they correspond to generators?
In Sec.~\ref{sec_ex1} we started with the description of a density in space, 
a volume smoothly filled with ``matter''. By the use of the Fourier transform , 
we switched to an {\it ensemble} of waves, the ``wave-packet''. By introducing
the wave-particle-duality (Eq.~\ref{eq_correspondence}) however, we 
introced a new Hamiltonian and by doing so we (implicitely) introduced an
ensemble of oscillators in some Hamiltonian phase space by Eq.~\ref{eq_dispersion}. 
We did not make that very explicite in Sec.~\ref{sec_ex2}, but here we explicitely 
consider (non-interacting) {\it ensembles} of solutions of Eq.~\ref{eq_eqom}.

Classical ensembles of non-interacting (or weakly interacting) systems
are subject of classical statistical mechanics, similar to ensembles of
particles in accelerator bunches and can be described by a phase space 
density $\rho(\psi)$. 
In contrast to ensembles from classical mechanics, where the density
is a density of a huge but countable number of ``mass points'', the density
we presume here is a smooth and continuous distribution in phase space. 
Distributions can of course be described by various mathematical methods. 
One possibility is a description based on the moments 
$\langle q_i^\mu\,p_j^\nu\rangle$ of the distribution.
The most important moments are the second moments, represented
by the (``auto-correlation'') matrix $\Sigma$ of second moments.
The autocorrelation matrix allows to construct the desired correspondence 
between observables and generators: There are ten independent parameters 
in the symmetric $4\times 4$ matrix ${\bf A}$ (and hence in the Hamiltonian 
matrix ${\bf F}$) and also ten parameters in the (symmetric) matrix $\Sigma$. 
Let $\Sigma=\langle\psi\,\psi^T\rangle$ be the 
matrix of second moments of solutions of Eq.~\ref{eq_eqom}, then it follows that
\myarray{
\dot\Sigma&=\langle\dot\psi\,\psi^T\rangle+\langle\psi\,\dot\psi^T\rangle\\
&={\bf F}\,\langle\psi\,\psi^T\rangle+\langle\psi\,\psi^T\rangle\,{\bf F}^T\\
&={\bf F}\,\Sigma+\Sigma\,{\bf F}^T\\
\label{eq_preHeisenberg}
}
Eq.~\ref{eq_preHeisenberg} is well-known in accelerator physics and used to 
describe the development of the second moments of a distribution of particles 
within a frame co-moving with the bunch. The second moments allow to 
define the RMS-``size'' of the beam by the diagonal elements 
$\Sigma_{11}=\langle x^2\rangle$\footnote{
Of course, in accelerator physics, the involved matrices are in general
of size $6\times 6$. As mentioned before, median plane symmetry often
reduces the size of the problem effectively to $4\times 4$ and $2\times 2$.}.

Accelerator physics is mostly satisfied with Eq.~\ref{eq_preHeisenberg}, but
let us take one more step which enables to arrive at a much more transparent 
framework\cite{Wolski} .
This step consists in by a multiplication of both sides of Eq.~\ref{eq_preHeisenberg}
with $\y_0^T$ from the right and in the definition of a Hamiltonian matrix 
${\bf S}=\Sigma\,\y_0^T$. Then one obtains the following equation of motion
for second moments:
\begeq
{\bf\dot S}={\bf F}\,{\bf S}-{\bf S}\,{\bf F}\\
\label{eq_Heisenberg}
\endeq
This equation is the general equation of motion of any Hamiltonian matrix
${\bf F}$. The general transformation law of these matrices is that of a
symplectic transformation:
\begeq
{\bf F}(\tau)={\bf M}(\tau){\bf F}(0){\bf M}^{-1}(\tau)
\endeq
with a symplectic matrix ${\bf M}(\tau)$, which can always be written
as a matrix exponent of another Hamiltonian matrix ${\bf G}$:
\begeq
{\bf M}(\tau)=\exp{({\bf G}\,\tau/2)}
\endeq
and
\begeq
{\bf M}^{-1}(\tau)=\exp{(-{\bf G}\,\tau/2)}={\bf M}(-\tau)
\endeq
Therefore one finds in few steps~\footnote{
  Whether or not the right side of Eq.~\ref{eq_Heisenberg2}
  has to be scaled by a factor of $1/2$ depends on the choice
  of units for $\tau$ or ${\bf G}$, respectively.
}:
\begeq
    {\bf\dot F}={d{\bf F}\over d\tau}=\frac{1}{2}\,({\bf G}\,{\bf F}(\tau)-{\bf F}(\tau)\,{\bf G})\,.
    \label{eq_Heisenberg2}
\endeq
This is Heisenberg's equation of motion for operators, an equation that
is usually regarded as intrinsically ``quantum''. It is automatically identical
to the equation of motion of observables, if the observables are second moments
in phase space.

The fact that quantum mechanics and (classical) beam dynamics are based on the
same mathematical (namely symplectic) structures has been recognized and
emphasized by many~\cite{DragtPRL,Forest,Dattoli1,Dattoli2,Qin2013,Gosson},
but is mostly ignored in common textbooks on quantum theory. Some
physicists believe that it is ``misleading'' to emphasize the common mathematical
basis of quantum and classical theory, namely Hamiltonian mechanics. But how 
can it be misleading, if not due to the presupposition that classical 
and quantum mechanics are fundamentally different? 

If it is not immediately obvious, that Eq.~\ref{eq_Heisenberg} is literally
Heisenberg's equation of motion, then likely because the quantum ``look and
feel'' requires the use of the unit imaginary and of Planck's constant $\hbar$. 
If we restrict our considerations to non-singular systems, then those 
eigenvalues of ${\bf F}$, that represent stable oscillation, are purely 
imaginary~\cite{MHO}. 
Furthermore they have the unit of a frequency. Hence, the 
``classical operator'' (i.e. the matrix) ${\bf H}$ has imaginary
eigenvalues, representing the frequencies of oscillation, and hence those 
of the operator $i\,{\bf H}$ are real.
When multiplied by $\hbar$, the real eigenvalues have the unit of energy
\begeq
{\bf H}=\mp i\,\hbar\,{\bf F}\,,
\endeq
so that one obtains:
\begeq
{\bf\dot S}=\pm i/\hbar\,({\bf H}\,{\bf S}-{\bf S}\,{\bf H})\,.
\label{eq_Heisenberg3}
\endeq
This testifies that much of the differences between equations appearing
in QM and those of classical Hamiltonian mechanics are due to 
the specific notation used in the former, which has been established 
in the course of the development, i.e. by convention.
But a successful description of nature does not depend on notational 
conventions and no system of equation becomes ``quantum'' just because 
we use the unit imaginary explicitely instead of implicitely. 
But while Bohr's complementarity and Heisenberg's uncertainty are supposed
to characterize the apparent fundamental difference between classical and
quantum notions, Dirac aimed for the opposite and usually emphasized
the analogy between Poission brackets and commutators~\cite{Dirac1963}:
\begin{quotation}
Hamilton made a further development of dynamical theory. He set up a
formalism which describes, not an individual solution of the equations of
motion, but a whole set of solutions together. [...]
At the time when Hamilton set up his theory there was no physical reason
why one should be interested in a family of solutions rather than an individual
Solution. The latter seemed quite adequate for a description of Nature.
Hamilton must have been inspired to know that his work was important.
He showed that there was an analogy between his dynamical theory and
geometrical opties, (namely that optics in which one neglects effects arising
from the finite wave-length of light.) But this analogy appeared at that
time to be just a mathematical curiosity without physical significance.
It needed a hundred years of progress in physics to show up the value of
Hamilton's work. Both of Hamilton's developments of dynamical theory are
of the greatest importance for quantum mechanics and are thus needed for a
description of Nature on the atomic scale.
[...]
The concept of the P. b. [Poisson bracket] is the all-important link
in the passage from classical to quantum mechanics, and this concept enters
into classical mechanics only with Hamilton's form of the theory.
\end{quotation}
However Eq.~\ref{eq_Heisenberg2} is a direct and {\it literal} consequence 
of Hamiltonian methods. Provided the Hamiltonian concerns ensembles of 
partial waves that inhabit an abstract phase space as suggested by the 
dispersion relation (Eq.~\ref{eq_dispersion}), then one has Dirac's
wave equation in momentum representation. Apparently it needs 
another century to admit that this is more than an ``analogy'', 
more than a ``mathematical curiosity without physical significance''. 

It follows from Eq.~\ref{eq_Heisenberg} that the second moments are
constants of motion when ${\bf\dot S}=0$, i.e. if ${\bf S}$ and ${\bf F}$ 
commute. It is a math fact of linear algebra that commuting matrices share a 
system of eigenvectors and that these eigenvectors must be complex, whenever
the eigenvalues are imaginary, i.e. in case of stable systems. 
${\bf S}$ provides the simplest possible (though in most cases incomplete)
description of phase space ensembles\footnote{Only if an ensemble is Gaussian, 
the matrix of second moments ${\bf S}$ provides a complete description.}.
Applying Eq.~\ref{eq_motion}, the autocorrelation matrix ${\bf S}(\tau)$ 
of the phase space ensemble as a function of time is given by
\myarray{
\Sigma(\tau)&=\langle\psi(\tau)\psi(\tau)^T\rangle\\
&={\bf M}(\tau)\langle\psi(0)\psi(0)^T\rangle\,{\bf M}^T(\tau)\\
&={\bf M}(\tau)\,\Sigma(0)\,{\bf M}^T(\tau)\\
}
This equation, at first sight, seems to suggest that the evolution
in time is an orthogonal transformation. However ${\bf M}$ is
orthogonal only in special cases, but always symplectic 
(Eq.~\ref{eq_symplectic}).
Again we multiply by $\y_0^T$ from the right and obtain
\begeq
{\bf S}(\tau)={\bf M}(\tau)\,{\bf S}(0)\,{\bf M}^{-1}(\tau)\,,
\endeq
where we used Eq.~\ref{eq_symplectic} in the last step:
The symplectic evolution in time is a similarity transformation, but
not necessarily an orthogonal one. 
Hence the eigenvalues of ${\bf S}$ are constants of motion\footnote{
They are ``emittances'', decorated by a unit imaginary~\cite{Wolski}.
In accelerator physics, the matrix ${\bf M}$ is the so-called
``transport matrix''. It is a product of the transport matrices of all
involved beam guiding elements (bending magnets, quadrupole magnets, 
buncher etc., see Ref.~\cite{rdm_paper}) and it is determined by 
the properties of the beamline elements, i.e. the ``outside world''. 
A beam described by ${\bf S}(\tau)$ is called ``matched'' to a given 
beamline described by ${\bf M}$, if ${\bf S}$ and ${\bf M}$ 
commute. This description is reasonably accurate as long as both, 
nonlinear terms and self-interaction by space charge or intra-beam-scattering 
can be neglected. However, if bunches have a non-negligible self-interaction
due to space charge, the matrix ${\bf F}$ and hence ${\bf M}$ also depends 
on the size of the beam: then ${\bf F}$ itself depends on (elements of)
${\bf S}$~\cite{sc_paper}.}.

There is a theorem in classical statistical mechanics about 
ensembles in phase space, which states that the constant phase space 
density of (thermal) equilibrium is a function of the energy, 
i.e. the Hamiltonian, or more generally, a function of the constants 
of motion, hence in our case, of the eigenvalues~\cite{Heat}. 
If $\Lambda$ is the (diagonal) matrix of eigenvalues of ${\bf M}$ and 
$\lambda$ the (diagonal) matrix of eigenvalues of ${\bf S}$, then, 
applied to the case at hand, this means that, in equilibrium, 
${\bf S}=f({\bf M})$ can be reduced to $\lambda=f(\Lambda)$. This is the 
case iff ${\bf S}$ and ${\bf M}$ have a common system of eigenvectors. 
Hence thermal equilibrium corresponds to a matched distribution and 
we can leave the question what exactly determines the form of 
${\bf F}$ open: both, external fields but also the properties of the 
considered system itself might be responsible for the precise form of ${\bf F}$. 

It is a known, though maybe not well-known, math fact that real 
$4\times 4$-matrices can be parameterized by the use of a Clifford 
algebra. Hestenes elaborated in detail how the Dirac Clifford algebra
generates geometrical significance~\cite{STA}. Insofar our approach
is, yet again, close to known presentations.
But here we use an approach slightly different from that of Hestenes.
We shall derive the Dirac algebra from Hamiltonian notions based on the 
SUM $\y_0$ as an essential structure generating element\footnote{A very brief 
intro to Clifford algebras is given in App.~\ref{sec_CA}.}.
In fact {\it any} real squared matrix of size $2^n\times 2^n$ can be 
written as a sum
\begeq
{\bf M}=\sum\limits_{k=0}^{15}\,m_k\,\y_k\,,
\endeq
where $\y_k$ are the unit elements of the Clifford algebra and the index $k$ 
runs over all unit elements (vectors, bi-vectors, etc.). 
But why should it be sensible to apply such a change of variables 
from profane matrix elements $m_{ij}$ to something fancy like the coefficients 
of a Clifford algebra? Is this necessary or just ornamental like $\hbar$
and the unit imaginary in QM? Can we make the case from the perspective 
of Hamiltonian mechanics?

The representation by Clifford algebras charges numbers ($m_k$) with 
``structural significance''. The simplest case of one degree of freedom 
requires only $2\times 2$ matrices:
\begeq
{\bf M}=\bmtx{cc}m_{11}&m_{12}\\m_{21}&m_{22}\emtx\,.
\endeq
But since a Hamiltonian matrix ${\bf F}=\y_0\,{\bf A}$ is a product 
of a skew-symmetric and a symmetric matrix, it has a vanishing trace. 
That is, Hamiltonian matrices have the boundary condition of a vanishing
trace, here $m_{11}+m_{22}=0$. Therefore we define new variables 
$c=m_{11}-m_{22}$ and $d=m_{11}+m_{22}$ and obtain
\begeq
{\bf M}=\bmtx{cc}(c+d)/2&m_{12}\\m_{21}&(d-c)/2\emtx\,,
\endeq
so that the parameter $d$ is directly proportional to the trace.
But as we have shown, also the distinction of symmetric and skew-symmetric 
elements is of severe importance in linear Hamiltonian theory,
so that eventually we write
\begeq
{\bf M}=\bmtx{cc}(c+d)/2&(a+b)/2\\(b-a)/2&(d-c)/2\emtx\,,
\endeq
and out pops the representation of a Clifford algebra, namely of
the real Pauli algebra $Cl(1,1)$ or $Cl(2,0)$, respectively\footnote{
The usual complex Pauli matrices are a reduction derived from the Dirac 
algebra and are {\it therefore} complex.
}:
\begeq
{\bf M}=a\,\eta_0+b\,\eta_1+c\,\eta_2+d\,{\bf 1}\,.
\label{eq_Pauli}
\endeq
A symmetric matrix corresponds to $a=0$, and
$d=0$ implies a matrix with vanishing trace: We thus constructed a 
scheme in which numbers (quantities) receive structural significance:
Quantity and structure become ``entangled'', but in a systematic way,
so that all coefficients $a,b,c,d$ quantify specific symmetries of the
Clifford algebra and hence of the dynamical properties of the system.

It is not directly evident from Eq.~\ref{eq_Pauli} that the derived set
of four matrices $\eta_k$ is indeed the representation (rep) of a 
Clifford algebra (CA), but it becomes evident if we look at the
anti-commutators:
\begeq
\eta_i\,\eta_j+\eta_j\,\eta_i=\pm\,2\,\delta_{ij}
\endeq
Hence all Pauli-matrices square to $\pm\,{\bf 1}$ and all
of them either commute of anti-commmute with all others. Then
they are a representation (rep) of some CA. There is no need to
{\it postulate} physical significance of CAs. The physical significance
of Clifford algebras can be obtained from Hamiltonian notions alone.
And indeed, quadratic forms and Clifford algebras are deeply related~\cite{Lounesto,Porteous}. 

The Hamiltonian symmetries introduced for $2\times 2$-matrices are 
preserved (and more emerge), if matrices of more complex systems with 
more degrees of freedom are constructed from the real Pauli algebra 
by Kronecker multiplication. For two degrees of freedom, we have to 
consider all Kronecker products of the (real) Pauli matrices 
(Eq.~\ref{eq_Pauli}) and out pops the real Dirac algebra. 
More generally it seems that any Clifford algebra that fully conforms 
with Hamiltonian notions, has a rep that can be obtained from
Kronecker products of the real Pauli algebra.

Since the Dirac algebra meets the symmetries of Hamiltonian 
mechanics, the distinction between Hamiltonian and skew-Hamiltonian 
elements splits the $16$ coefficients $m_k$ into two sets of matrices, 
$10$ of which are Hamiltonian and $6$ skew-Hamiltonian.
A Hamiltonian $4\times 4$ matrix can then be written as
\begeq
{\bf F}=\sum\limits_{k=0}^{9}\,f_k\,\y_k\,.
\label{eq_HamiltonioanMatrix}
\endeq
We have shown that the use of Clifford algebras in Hamiltonian theory 
can be motivated purely by Hamiltonian symmetries, but one can only
make use of the Clifford algebraic approach, if the matrix ${\bf A}$ is 
of size $2^n\times 2^n$, for instance in our case of the coupling of
two degrees of freedom.

The Fourier transformation used in Sec.~\ref{sec_ex2} is a (unitary)
transformation to new variables, and the use of the real Dirac algebra 
is but another transformation to new variables. It is a general phenomenon 
that most work to solve a (solvable) physical problem is done when we 
have found the transformation to appropriate variables. Though the use 
of a Clifford algebra results from the analysis of dynamical symmetries 
of pure classical phase space, it nonetheless is a new element that
was unknown in classical pre-quantum physics. The reason is that this is 
a method of maximal generality. Before the advent of quantum mechanics,
``classical'' mechanics was mostly used to describe specific
systems with specific forms of ${\bf F}$. And though ensembles in phase 
space were subject of statistical mechanics, it was mostly understood 
as the phase space of ensembles of point-particles in 3d-space (and time).

As we shall briefly sketch in the following, the Dirac algebra has the 
additional and unexpected feature to automatically provide us with a 
unique interpretation in the sense, that the commutation properties of 
the algebra alone suffice to determine the transformation properties 
of all Hamiltonian coefficients~\cite{qed_paper,osc_paper}. 
This automatically and inevitably generates a system of $10$ variables
and their behavior under symplectic transformations (generated by the
same $10$ quantities) which is able to represent the well-known set of 
physical quantities which are relevant for the description of a relativistic
charged particle in an external electromagnetic field\footnote{
A detailed demonstration of the inevitability exceeds the scope of
this paper, but has been given in Ref.~\cite{qed_paper,osc_paper}.},
namely energy ${\cal E}$ and the cartesian components of momentum
${\vec P}$, electric and magnetic field $\vec E$ and $\vec B$, respectively. 

The analysis of the elements of the Clifford algebra that is represented 
by real $4\times 4$ matrices naturally begins with the distinction between 
Hamiltonian and skew-Hamiltonian matrix elements.
It follows from Eq.~\ref{eq_symplex} that $\y_0$ itself is Hamiltonian.
It is therefore the first of $10$ parameters necessary to specify the
linear Hamiltonian system of two canonical pairs.
If $\y_0$ is the first {\it basis} element of the Clifford algebra, then any
other {\it basis} element $\y_a$ must anti-commute with $\y_0$. This
follows from the definition of Clifford algebras. 
If one furthermore demands that all basis elements $\y_a$ must be 
Hamiltonian, then all other basis elements, except $\y_0$, must be symmetric 
(see Eq.~\ref{eq_ac_sym}) and therefore square to $+{\bf 1}$. 
We call a real Clifford algebra (CA) with purely Hamiltonian basis a 
{\it Hamiltonian} Clifford algebra (HCA).
As we just derived and explained, {\it any} Hamiltonian Clifford algebra 
of dimension $N=p+q$ in which the SUM $\y_0$ is a generating element has 
dimension $Cl(N-1,1)$ and therefore produces a metric of Minkowski type. 

Real $4\times 4$ matrices may represent either $Cl(2,2)$ or $Cl(3,1)$, each 
having $4$ basis elements, but only $Cl(3,1)$ is Hamiltonian, i.e. has
a Clifford basis $\y_\mu$ consisting exclusively Hamiltonian 
elements\footnote{This is required in order to have a
``dimension'' that is able to act as a Hamiltonian generator. See below.}. 

Dirac introduced $4\times 4$ matrices in order to {\it reproduce} the already
known RDR ${\cal E}^2-\vec p^2=\mathrm{const}=m^2$ (using $c=1$). But no 
other Dirac Clifford algebra but only $Cl(3,1)$ is based exclusively
on (real) Hamiltonian matrices. In the conventional, historically oriented
lore, Lorentz covariance is a more or less surprizing requirement for the
invariance of Maxwell's equations, which have been discovered experimentally 
and combined piece by piece by Faraday, Maxwell, Heaviside and others. Then 
it was Einstein's principle of the constancy of the speed of light, that 
required to establish the Lorentz transformations as something fundamental\cite{Dirac80}:
\begin{quotation}
The real importance of Einstein's work was that he introduced
Lorentz transformations as something fundamental in physics.
\end{quotation}

However, there are always different perspectives possible, 
as described by Swann~\cite{Swann}:
\begin{quotation}
In any presentation of a branch of modern physics, two courses are open.
The first is the historical. This has the disadvantage that, usually, it does
not represent a sequence of logical developments. The ways in which
conclusions are reached are founded frequently upon considerations of special
cases, and sometimes are based upon experiments whose representatives
have completely evaporated in the more general fields in which the conclusions
are subsequently used. The alternative method is to take the results
which have been stumbled upon in the historical development, review the
path by which they have been reached, remove from it as much as possible
of the obsolete debris with which it is encumbered, and try to construct some
more satisfactory path by which the results might have been reached even to
the extent of possible modifications in the starting points.
\end{quotation}
or by Peter Ball in his recently published brilliant book\cite{Ball}:
\begin{quotation}
Yet most popular descriptions of quantum theory have
been too wedded to its historical evolution. There is no
reason to believe that the most important aspects of the
theory are those that were discovered first, and plenty of
reason to think that they are not.
\end{quotation}
Levy-Leblond summarized it even shorter\cite{LevyLeblond}:
\begin{quotation}
The chronological building of order of a physical theory, however, 
rarely coincides with its logical structure.
\end{quotation}
Though this could be regarded as a platitude, the lesson apparently
has not yet been understood. Why else should it be emphasized?

Nothing in the usual presentation of the matter suggests, that the RDR 
can, with the help of the Clifford algebras, be obtained from 
classical Hamiltonian notions.
But it is a math fact, that the Clifford algebraic structure $Cl(3,1)$ 
stems from the symmetry of classical phase space~\cite{qed_paper,osc_paper} 
and this suffices to derive the mathematical form of the Lorentz 
transformations. But then the core concepts of the physics of the 20th 
century, namely Lorentz covariance and wave mechanics, are barely more 
than applied Hamiltonian mechanics
\footnote{We are aware of the fact that the mere form of equations 
  does not automatically provide an interpretation. The method suggested
  by Loewdin's contentless deductive theory implies that possible
  interpretations (and hence applications) have to be found on the
  basis of the formal relations. See also the discussion
  in Ref.~\cite{fit_paper}.}.

But the usual textbook presentations of special relativity discusses 
coordinate transformations as something physical without recurring to
the physical quantities which are the generators of these transformations\footnote{
Brent Mundy made a related remark~\cite{Mundy1986}:
\bquo
There are several respects in which the standard formulation may be
considered as inadequate or misleading, from a philosophical viewpoint.
In the first place, it leaves some uncertainty as to what the theory is a
theory {\it of}. Taking the standard presentation literally, it seems to be
a theory of coordinate systems and their properties and relations. This
is somewhat disturbing, since a coordinate system is, after all, an arbitrary
and artificial human construct, part of our conceptual apparatus for the
{\it description} of nature, rather than a proper part of the subject matter
of physics itself.
\equo
}.
It is part of Hamiltonian methods to regard physically (and not merely
mathematically) possible transformations as generated by observable physical 
quantities. The saying that the Hamiltonian function itself is ``the generator of 
translation in time'', expresses the content of Eq.~\ref{eq_motion}.
As we shall demonstrate in Sec.~\ref{sec_ex4} and Sec.~\ref{sec_ex5}, 
the generators of both, rotations and boosts, can be identified with physical 
observables, namely the magnetic and electric fields, respectively\footnote{ 
However, electric and magnetic field are the generators of boosts and rotations 
not in coordinate (``physical'') space, but in {\it energy-momentum-space}, 
as immediately obvious from the mathematical form of the Lorentz force.}.
In the last example in Sec.~\ref{sec_ex5} it will be shown that even 
Maxwell's equations can be obtained from Hamiltonian 
considerations~\cite{qed_paper}.

We introduced the notion of the Hamiltonian Clifford basis, from which all 
other elements of a Clifford algebra are generated. All basis elements
combined give a four-parameter matrix ${\bf F}$ with the form\footnote{The 
explicite form is given in Eq.~\ref{eq_Pmtx}.}:
\begeq
{\bf F}=\w\,\y_0+k_1\,\y_1+k_2\,\y_2+k_3\,\y_3\,,
\label{eq_Fvec}
\endeq
where $\y_0^2=-{\bf 1}$ and $\y_k^2={\bf 1}$ with $k=[1,2,3]$ are mutually
anti-commuting Hamiltonian matrices. Using only these basic elements from
which the Clifford algebra is generated, the equations of motion 
(Eq.~\ref{eq_eqom}) have the general form
\begeq
\dot\psi=(\w\,\y_0+k_1\,\y_1+k_2\,\y_2+k_3\,\y_3)\,\psi
\endeq
so that we obtain a ``2-dimensional'' stable oscillator
\begeq
\ddot\psi=(\w\,\y_0+k_1\,\y_1+k_2\,\y_2+k_3\,\y_3)^2\,\psi=-\w_0^2\,\psi
\endeq
with the invariant eigenfrequency $\w_0^2=\w^2-k_1^1-k_2^2-k_3^2$ for $\w_0^2>0$. 
This allows to obtain a purely Hamiltonian dispersion relation and as a matter 
of fact it is the correct relativistic dispersion relation (RDR). The only 
remaining step is to show that the time variable $\tau$ in the time derivative
$\dot\psi={d\over d\tau}\psi$ is not the laboratory but indeed the proper
time. Then, with the de Broglie relations from above\footnote{
We shall come back to this in part two.}, one obtains
\begeq
\hbar\,\dot\psi={\bf F}\,\psi=({\cal E}\,\y_0+p_1\,\y_1+p_2\,\y_2+p_3\,\y_3)\,\psi
\label{eq_Fvec1}
\endeq
so that the mass $m=\sqrt{{\cal E}^2-\vec p^2}$ is both, an invariant 
eigenvalue of ${\bf F}$, but also a constant of motion. However,
it is a constant of the motion of $\psi$, which can not be observed 
directly in a classical sense
(we come back to this in Sec.~\ref{sec_hamilton}). To describe the
motion of the unobservable quantities $\psi$ is of limited physical
value. It is therefore required to change the dynamical variables
and to introduce a new Hamiltonian that depends on the observables
${\cal E}$ and $\vec p$, i.e. on the second moments ${\bf S}$. This 
step converts the status of the mass, the value of the first Hamiltonian, 
into a mere invariant parameter\footnote{This kind of flexibility to 
chose Hamiltonians is well established in the kind of classical 
mechanics developed for accelerators~\cite{Guignard}.}. 

As we shall show below (Eq.~\ref{eq_LT}), skew-symmetric Hamiltonian generators 
yield rotations and symmetric ones generate boosts.
Since (as shown above) all skew-symmetric Hamiltonian generators\footnote{
It will be shown below that skew-symmetric matrices generate rotations
while symmetric matrices generate boosts.} commute
with $\y_0$, they can't change the value of ${\cal E}$ (see Eq.~\ref{eq_trans}). 
Hence ${\cal E}$ is the only rotationally invariant vector component known so 
far, and it is therefore nearby to use it as next Hamiltonian function. 
The canonical conjugate of the energy ${\cal E}$ is a new time coordinate
$t$. The relation between old ($\tau$) and new time variable $t$ 
follows from ${\cal H}=m\,c^2=\sqrt{{\cal E}^2-\vec p^2\,c^2}$:
\myarray{
{dt\over d\tau}&=\dot t={\d{\cal H}\over\d{\cal E}}\\
&={{\cal E}\over{\cal H}}={{\cal E}\over m\,c^2}
\label{eq_timedilation}
}
We now use the ``quantization rules'' (Eq.~\ref{eq_cquant}) to replace the 
total derivative on the left side of Eq.~\ref{eq_Fvec1} by the corresponding 
partial derivatives on the right to obtain the Dirac equation in the usual notation
\myarray{
-i\,m\,c^2\,\psi&=i\,\hbar\,(\d_t\,\y_0-\d_1\,\y_1-\d_2\,\y_2-\d_3\,\y_3)\,\psi\\
m\,c^2\,\psi&=i\,\hbar\,(\d_t\,\G_0-\d_1\,\G_1-\d_2\,\G_2-\d_3\,\G_3)\,\psi\\
\label{eq_imaginary}
}
where $\G_\mu=i\,\y_u$ are the conventional complex Dirac matrices corresponding to
the conventional metric tensor $g_{\mu\nu}=\mathrm{Diag}(1,-1,-1,-1)$. 
Hence the unit imaginary is, within our approach, an artifact of the preference 
for the metric $g_{\mu\nu}=\mathrm{Diag}(1,-1,-1,-1)$ instead of the use of a metric
$g_{\mu\nu}=\mathrm{Diag}(-1,1,1,1)$: The use of the unit imaginary in the
Dirac equation is an excercise in redundancy. 

It is mostly agreed that the sign of the metric has no physical 
significance\footnote{
Most textbooks use the metric $g_{\mu\nu}=\mathrm{Diag}(1,-1,-1,-1)$. 
Weinberg's books on quantum field theory however uses 
$g_{\mu\nu}=\mathrm{Diag}(-1,1,1,1)$~\cite{WeinbergQFT}.
}. However, the conventional metric leads to a notation that suggests
that the unit imaginary is a meaningful and necessary ingredient in 
Dirac's theory, something that generates ``quantumness''. But as we demonstrated,
Dirac's theory allows for, but neither suggests nor requires the explicit
use of the unit imaginary\footnote{Since Schr\"odinger's original equation 
does not use spinors, the wave function must be complex in order to provide 
a canonical pair~\cite{Strocchi,Ralston1989,osc_paper,uqm_paper}.}.

Coming back to the ``particle picture'' one obtains the new Hamiltonian
dispersion relation ${\cal H}(\vec p)$, in the new time coordinate $t$, 
which then reads 
\begeq
{\cal H}=\sqrt{m^2\,c^4+\vec p^2\,c^2}\,,
\endeq
which results in the Hamiltonian velocity of a free particle 
(Eq.~\ref{eq_someHamiltonian}):
\begeq
\vec v=\vec\nabla_p\,{\cal H}(\vec p)={\vec p\,c^2\over\sqrt{m^2\,c^4+\vec
    p^2\,c^2}}={\vec p\,c^2\over{\cal E}}\,.
\label{eq_velo0}
\endeq
where the velocity is, using the new Hamiltonian, the temporal
derivative with respect to the coordinate time $t$ (and not $\tau$):
\begeq
\vec v={d\vec x\over dt}\,.
\endeq
If we scale to the constant $c$, then this reads as
\begeq
\vec\beta={\vec v\over c}={\vec p\,c\over{\cal E}}\,.
\label{eq_velo1}
\endeq
Solving for ${\cal E}$ and ${\vec p}$, one readily obtains
\myarray{
{\cal E}&=m\,c^2\,\y\\
\vec p&=m\,c\,\y\,\vec\beta\\
\label{eq_relpE}
}
using the usual definition of $\y={1\over\sqrt{1-\vec\beta^2}}$.
Combining Eq.~\ref{eq_timedilation} and Eq.~\ref{eq_relpE}, we obtain
``time dilation'' $dt=\y\,d\tau$ as a result of a canonical transformation. 
In a preceeding paper we elaborated in detail that the Lorentz 
transformations are canonical symplectic similarity transformations 
and have their (conceptually) simplest representation in the $4\times 4$ 
real Dirac algebra~\cite{lt_paper}. In the next section we will sketch 
the general setting.

As well-known, one arrives at the Newtonian expression in the usual
approximation, taking only the first terms of the Taylor serie 
of ${\cal E}(\vec p)$:
\begeq
{\cal E}=m\,c^2+{\vec p^2\over 2\,m}+\dots
\endeq
which yields, due to $\vec v=\vec\nabla_p(E)$ Newton's $\vec p=m\,\vec v$.
Furthermore the theory defines, what may and what may not be constant.
If ${\bf S}$ and ${\bf F}$ commute, then both ${\cal E}$ and $\vec p$
and hence the velocity $\vec v$ is constant. This, in some sense,
(re-) establishes Newton's first axiom -- from a Hamiltonian point of view.

\section{Fourth Example: Lorentz Transformations}
\label{sec_ex4}

It is well known and understood from the theory of Lie algebras that 
Hamiltonian observables are generators of canonical transformations.
Usually, when we employ a Hamiltonian description of a system of 
classical oscillators, our (macroscopic) description of the involved
masses and spring constants determines the exact form of the matrix 
${\bf F}$, i.e. which of the $10$ possible parameters of ${\bf F}$
vanish, which do not, and to what physical quantity they are related. 
But since we aim for the most general description, we have no 
reason to assume that certain elements of ${\bf F}$ have some
specific value. Since there are $10$ free parameters in ${\bf F}$
in total, six parameters are left to be discussed. 

These $6$ parameters can be devided into two groups, firstly a set 
of three symmetric matrices
\myarray{
\y_4&=\y_0\,\y_1\\
\y_5&=\y_0\,\y_2\\
\y_6&=\y_0\,\y_3\\
\label{eq_elec}
}
and secondly a set of three skew-symmetric matrices:
\myarray{
\y_7&=\y_{14}\,\y_0\,\y_1=\y_2\,\y_3\\
\y_8&=\y_{14}\,\y_0\,\y_2=\y_3\,\y_1\\
\y_9&=\y_{14}\,\y_0\,\y_3=\y_1\,\y_2\,.
\label{eq_mag}
}
It is a math fact that bi-vectors, products of two Hamiltonian
basis elements $\y_\nu$, are also Hamiltonian, while
$3$-vectors and $4$-vectors are skew-Hamiltonian~\cite{rdm_paper,geo_paper,qed_paper,osc_paper}. 
Therefore the $6$ missing Hamiltonian parameters come in two
sets of $3$ bi-vector elements each. Note that this grouping 
into $3$-vectors results from Hamiltonian symmetries.

If we consider the general properties of transformations using
Eq.~\ref{eq_motion} with single Hamiltonian Clifford elements $\y_a$
for which $\y_a^2=\pm{\bf 1}$:
\begary{rclp{3mm}rcr}
{\bf M}_a(\tau)&=&{\bf 1}\,\cos{(\tau)}+\y_a\,\sin{(\tau)}&&\mathrm{if}&\y_a^2=&-{\bf 1}\\
               &=&{\bf 1}\,\cosh{(\tau)}+\y_a\,\sinh{(\tau)}&&\mathrm{if}&\y_a^2=&{\bf 1}\\
\label{eq_LT}
\endary
Note that ${\bf M}^{-1}(\tau)={\bf M}(-\tau)$ holds for all
transformations of Eq.~\ref{eq_LT}. 
Whether such a transformation leaves some element constant
or not, depends exclusively on the commutation properties of the
algebra. Since the transformation matrices for pure transformations
Eq.~\ref{eq_LT} contain only ${\bf 1}$ and $\y_a$, they commute
with some $\y_b$ exactly, if $\y_a$ and $\y_b$ commute. Then the
coefficient of $\y_b$ remains unchanged by the similarity transformation
\myarray{
\tilde\y_b&={\bf M}_a\,\y_b\,{\bf M}_a^{-1}\\
\tilde\y_b&=\y_b\,.
}
If $\y_a$ and $\y_b$ anti-commute ($\y_a\,\y_b=-\y_b\,\y_a$), 
however, we obtain (rotations, $\y_a^2=-{\bf 1}$):
\myarray{
\tilde\y_b&={\bf M}_a(\tau/2)\,\y_b\,{\bf M}_a^{-1}(\tau/2)\\
&=({\bf 1}\,\cos{(\tau/2)}+\y_a\,\sin{(\tau/2)})\,\y_b\\
&\times\,({\bf 1}\,\cos{(\tau/2)}-\y_a\,\sin{(\tau/2)})\\
&=(\cos^2{(\tau/2)}-\sin^2{(\tau/2)})\,\y_b\\
&-2\,\sin{(\tau/2)}\,\cos{(\tau/2)}\,\y_b\,\y_a\\
&=\cos{(\tau)}\,\y_b-\sin{(\tau)}\,\y_b\,\y_a\\
\label{eq_trans}
}
and boosts, correspondingly, for $\y_a^2={\bf 1}$~\cite{lt_paper}:
\myarray{
\tilde\y_b&=\cosh{(\tau)}\,\y_b-\sinh{(\tau)}\,\y_b\,\y_a\\
}
Hence any symplectic similarity transformation with pure Clifford
elements results in a rotation in phase space for skew-symmetric
matrices $\y_a^2=-{\bf 1}$ and in a boost for symmetric matrices 
$\y_a^2={\bf 1}$. Other, polynomial solutions are also possible,
but they do not represent non-singular systems and we do not adress 
them here~\cite{exp_paper}.

Many textbooks on QED do not elaborate the Lorentz transformation
of Dirac spinors in detail\footnote{
The best presentation known to the author, albeit in German, can
be found in Schm\"user's book~\cite{Schmueser}.}. We therefore 
refer to a preceeding paper 
in which we explicitely elaborated the Lorentz transformations
on the basis of these Hamiltonian notions~\cite{lt_paper}. 
It is both, a result of these investigations, but also well-known
in Dirac's theory that the components of the symmetric bi-vector
are generators of boosts and transform like the electric field,
i.e. like a so-called ``radial'' bi-vector $E_x\,\y_4+E_y\,\y_5+E_z\,\y_6$.
The components of the skew-symmetic ``axial'' bi-vector are generators of 
rotations and transform like the components of the magnetic field 
vector ${\vec B}=B_x\,\y_7+B_y\,\y_8+B_z\,\y_9$.

Hence there is another matrix ${\bf F}$, which consists of electromagnetic 
bi-vector components\footnote{The explicite form is given in Eq.~\ref{eq_Fmtx}.}:
\begeq
{\bf F}=E_x\,\y_4+E_y\,\y_5+E_z\,\y_6+B_x\,\y_7+B_y\,\y_8+B_z\,\y_9\,.
\label{eq_Fbivec}
\endeq
The eigenfrequencies of this matrix are the known relativistic invariants
\begeq
\w=\pm\sqrt{\vec B^2-\vec E^2\pm\,2\,\sqrt{-(\vec E\cdot\vec B)^2}}\,.
\label{eq_omega2}
\endeq
Of course, this equation makes only sense, if we can express fields
in units of frequencies. But the required physical scaling constants
are known today and effectively this means little more than to express
the electromagnetic fields in units of Schwinger's limiting 
fields~\cite{uqm_paper}.

The representation of structure by numbers as implemented by the
use of the Dirac algebra {\it automatically} delivers the most compact 
form of the Lorentz transformations~\cite{lt_paper}, but also the 
invariants of electromagnetic fields, even before we derived or
even considered Maxwell's equations at all.

Then it should not be surprising that also the Lorentz force
and Maxwell's equations pop out~\cite{qed_paper}. In order to
better distinguish vector components (Eq.~\ref{eq_Fvec}) 
from the bi-vectors components (Eq.~\ref{eq_Fbivec}) and the total
Hamiltonian matrix, we use a bold ${\bf P}$ for the 4-momentum:
\begeq
{\bf P}={\cal E}\,\y_0+p_x\,\y_1+p_y\,\y_2+p_z\,\y_3
\label{eq_P}
\endeq
and ${q\over m}\,{\bf F}$ for the bi-vectors (Eq.~\ref{eq_Fbivec}).
The factor ${q\over m}$ enters to obtain the equations in the usual
system of units (see also Ref.~\cite{uqm_paper}).
Then Eq.~\ref{eq_Heisenberg2} can be written as follows:
\begeq
{\bf\dot P}={q\over 2\,m}\,({\bf F}\,{\bf P}-{\bf P}\,{\bf F})\,.
\label{eq_lorentzforce}
\endeq
Written explicitely in vector components we have~\cite{rdm_paper,qed_paper,osc_paper}:
\myarray{
{d{\cal E}\over d\tau}&={q\over m}\,\vec p\cdot\vec E\\
{d\vec p\over d\tau}&={q\over m}\,\left({\cal E}\,\vec E+\vec p\times\vec B\right)\\
}
Using the lab frame time $dt=\y\,d\tau$ these equations
are identical to the usual Lorentz force equations (for $c=1$).
Hence also the Lorentz force can be obtained purely on Hamiltonian 
grounds, even without knowledge of Maxwell's equations.

\section{Last Example: Maxwell's Equations}
\label{sec_ex5}

Eq.~\ref{eq_Heisenberg2} has another important implication: The
change of a Hamiltonian of the left is side is connected to a 
product, namely the skew-symmetric product, of two Hamiltonian 
matrices on the right side. This is important, because it connects
the time evolution of $k$-vectors with $k\pm j$-vectors by a
multiplication.

We call a Hamiltonian Clifford algebra irreducible, if the 
maximal number of variables in the matrix representation $2\,n\times 2\,n=4\,n^2$
correponds to the number of elements of the Clifford algebra,
which is $2^N$. Equating these numbers $4\,n^2=2^N$ provides 
evidence that all irreducible Hamiltonian Clifford algebras have 
an even dimension $N$. In $p+q=N$ is even, then it is impossible 
to obtain all elements from bi-vectors only. No multiplication 
of any number of even elements may produce odd elements 
(vectors, $3$-vectors).
If in Eq.~\ref{eq_Heisenberg}, both ${\bf F}$ and ${\bf S}$, are
even and Hamiltonian, i.e. bi-vectors, then the left side is either 
a scalar or a pseudoscalar or another bi-vector. It can not be 
a vector.

Interpreting Eq.~\ref{eq_Heisenberg} physically, we can construct
bi-vectors from the interaction of vector quantities but not 
vice-versa. Hence we may regard vectors as representations 
of particles and bi-vectors as representations of fields, 
generated by particles.
Bi-vectors are the generators of rotations and boosts of vectors, 
but they can not directly be used to establish vectors by any kind 
of Lorentz covariant multiplication as in Eq.~\ref{eq_Heisenberg}.

It is part of Hamiltonian theory to distinguish mechanical
and canonical momentum. The (possible) difference appeared before in
Eq.~\ref{eq_mech_cano}: The relation between velocity and momentum
allows for additional components $\vec A$; correspondingly the energy
may contain an additional term $\phi$. When established by 
Eq.~\ref{eq_mech_cano}, then we have to consider an additional vector 
type quantity $\phi,\vec A$ that depends on coordinates only. 
It follows that we must in general regard those quantities that
do not depend on the momentum, i.e. the bi-vector coefficients, 
as dependent on the corresponding canonical coordinates:
\myarray{
\vec E&=\vec E(\vec x,t)\\
\vec B&=\vec B(\vec x,t)\\
}
and we should expect that these components can be obtained from 
vector type quantities $\phi$ and $\vec A$.

Again, as in the first two examples, the Hamiltonian method allows to 
derive equations of motion for new variables, this time for the 
Maxwellian bi-vector fields. 
First we need a derivative operator that is compatible with 
the Hamiltonian-Clifford framework elaborated so far: It must allow
for the described symplectic similarity transformation. 
The derivative operator is, of course, a vector type quantity:
\begeq
\d\equiv -\d_t\,\y_0+\d_x\,\y_1+\d_y\,\y_2+\d_z\,\y_3\,.
\label{eq_deriv}
\endeq
As established by Eq.~\ref{eq_Heisenberg}, Hamiltonian motion is connected 
to symmetric products (anti-commutators) and skew-symmetric products. 
Then matrix multiplication from the right combined with a derivative
$\d$ requires to indicate the direction in which the differentiation 
acts. We indicate the direction by arrows in what follows.
The commutative derivative is
\begeq
\d\wedge{\bf A}\equiv\frac{1}{2}\,\left(\rightD{\d}{\bf A}-{\bf A}\leftD{\d}\right)
\endeq
and the {\it anti-commutative} 
\begeq
\d\cdot{\bf A}\equiv\frac{1}{2}\,\left(\rightD{\d}{\bf A}+{\bf A}\leftD{\d}\right)
\endeq
Four different derivatives are possible with following results:
\begary{rcl}
\d\,\wedge\,\textrm{ vector }&\Rightarrow&\textrm{ bi-vector}\\
\d\,\wedge\,\textrm{ bi-vector }&\Rightarrow&\textrm{ vector}\\
\d\,\cdot\,\textrm{ vector }&\Rightarrow&\textrm{ scalar}=0\\
\d\,\cdot\,\textrm{ bi-vector }&\Rightarrow&\textrm{ axial vector}=0\\
\label{eq_difftypes}
\endary
There is only one unique way to express bi-vector fields from 
such a derivative -- it is the commutative derivative of a four 
vector, according to the first of Eq.~\ref{eq_difftypes}. 
This demonstrates the rigidity of Hamiltonian notions. We may
now write this equation, using the vector type ``potential'' 
${\bf A}=\y_0\,\phi+\vec\y\,\vec A$
\begeq
{\bf F}=\d\,\wedge\,{\bf A}\,,
\label{eq_pot_grad}
\endeq
or explicitely in components:
\begary{rcl}
\vec E&=&-\vec\nabla\phi-\d_t\vec A\\
\vec B&=&\vec\nabla\times\vec A\,.
\endary
This is the only possible linear Hamiltonian definition of the 
electromagnetic field from vector type quantities and it explains
the meaning of the ``integration constants'' appearing in
Eq.~\ref{eq_mech_cano}.

The second of Eq.~\ref{eq_difftypes} suggests that the ``source'' of
a bi-vector field is again a vector:
\begeq
\d\,\wedge\,{\bf F}=4\,\pi\,{\bf J}\,,
\label{eq_MWinhomo}
\endeq
which can be regarded as a {\it definition} of the {\it vector current}
\begeq
{\bf J}=\rho\,\y_0+j_x\,\y_1+j_y\,\y_2+j_z\,\y_3\,.
\endeq
Written explicitely in components, Eq.~\ref{eq_MWinhomo} is given by\footnote{
We have shown in Ref.~\cite{uqm_paper} that these equations are compatible
with the Dirac current.}:
\myarray{
\vec\nabla\cdot\vec E&=4\,\pi\,\rho\\
\vec\nabla\times\vec B-\d_t\vec E&=4\,\pi\,\vec j\,.
}

The third of Eq.~\ref{eq_difftypes} then yields the continuity equation
and likewise the Lorentz gauge. It is a trivial consequence of Eq.~\ref{eq_MWinhomo}:
\begeq
\d\,\cdot\,{\bf J}=\frac{1}{16\,\pi}\,\left(
\rightD{\d}^2{\bf F}-\rightD{\d}{\bf F}\leftD{\d}+\rightD{\d}{\bf F}\leftD{\d}-{\bf F}\leftD{\d}^2\right)=0\,.
\label{eq_continuity}
\endeq
Note that $\leftD{\d}^2$ and $\rightD{\d}^2$ are scalars (d'Alembert's
operator $\square=\vec\nabla^2-\d_t^2$ 
and hence $\rightD{\d}^2{\bf F}-{\bf F}\leftD{\d}^2=0$.
Written in components, Eq.~\ref{eq_continuity} is equal to
\begeq
\d_t\rho+\vec\nabla\cdot\vec j=0\,.
\endeq
Finally, the last of Eq.~\ref{eq_difftypes} gives
\begeq
\d\,\cdot\,{\bf F}=0
\endeq
which are the homogeneous Maxwell equations, when written in components:
\myarray{
\vec\nabla\cdot\vec B&=0\\
\vec\nabla\times\vec E+\d_t\vec B&=0\,.
\label{eq_MWhomo}
}
From a rigorous Hamiltonian point of view, this is the proper way 
to establish Maxwell's equations, namely a way that {\it inherently
implies} the nature of their ``covariance''. 

Note that neither the autocorrelation 
matrix ${\bf S}$ nor the Hamiltonian matrix ${\bf F}$ may contain non-zero
coefficients for the skew-Hamiltonian elements of the Dirac algebra,
i.e. for the scalar $\y_{15}\equiv {\bf 1}$, pseudo-scalar $\y_{14}$
and the axial vector components $\y_{14}\,\y_\nu$. 
Hence we must demand that the corresponding derivatives vanish (in the 
linear approximation we discuss here) as indicated in Eq.~\ref{eq_difftypes}. 
But as we have seen, this comes out automatically from the formalism as
a consequence of the fact that the space-time-derivative must be a vector
in the Hamiltonian Clifford Algebra $Cl(3,1)$.

\section{Aftermath}
\label{sec_aftermath}

The usual mind-set of modern physics as to be found in standard textbooks
suggests that theorizing in fundamental physics starts with the presumption 
of some background space-time, some kind of mathematical space, often 
equipped with fancy mathematical features\footnote{
Despite the fact that the majority of physicists is said to have accepted 
the Copenhagen interpretation of QM, which asserts that at the atomic level
physical processes can not be described as being objective in space and time.}.
Minkowski's space-time is such a background and it (re-) produces the 
mathematical feature of Lorentz covariance. Mathematically there is 
nothing wrong with this. But this mode of thinking is, from a logical 
point of view, disturbing: In textbooks on QM it is asserted that the ``majority''
of physicists has accepted the Copenhagen interpretation\footnote{
Despite the fact that the real content of this interpretation is still 
controversial.} of QM, which asserts 
that at the atomic level physical processes can not be described as happening
objectively in space and time. But on the next page, the conventional lore
insists that, on the fundamental level, there must be a space-time with 
unexplainable (``given'') physical properties. This excludes the principle 
possibility that the dimensionality of space-time has itself a {\it dynamical}
reason. 

We also started with the assumption of a Newtonian space-time in the second 
example: we presumed some Euklidean space-time and a distributed amount
of matter in it, described by a (normalizable) density distribution.
In the third example however, we addressed the question whether we can 
derive some general kind of dispersion relation from nothing but 
Hamiltonian (i.e. dynamical) notions. Once the idea to consider the space
of possible linear canonical transformations with two abstract classical
canonical pairs of dynamical variables -- as suggested by Dirac -- is
considered, the structure of spacetime as described by special relativity,
follows with logical necessity. Therefore the Hamiltonian approach described
here is based on a different kind of fundamental background, which has the
form of a (symplectic) phase space, similar to the $\Gamma$-space of classical 
statistical mechanics. However, this phase space received, in the course
of reasoning, a new kind of ontological significance. In other cases,
like accelerator physics, the terms that can possibly be found in the 
matrix ${\bf F}$, can be (and actually are) derived from spelling out the
consequences of classical $3$-dimensional reasoning. However, in the
presented approach the logic needs to be reversed: All {\it mathematically
  possible} terms of the abstract phase space are considered on equal
footing. Nonetheless they allow to derive the dynamical symmetries and
relations that hold in a space of higher dimension, namely in Minkowski
space-time.

In the historical presentation of special relativity, the lack of
a physical/logical legitimization of the Lorentz transformations left room
for quite a number of alternative ``space-time'' transformations\footnote{
See Ref.~\cite{Brown} and references therein.}. Per-Olov L\"owdin has shown, 
that few general and reasonable assumptions about space suffice to constrain
the possibilities to two forms of space-time transformations, namely those of 
Galileo and Lorentz~\cite{Loewdin98}. 
But the Hamiltonian framework that we described is even more restrictive 
and does not require the assumption of space-time at all. Furthermore it 
incorporates the Hamiltonian viewpoint that physically possible coordinate 
transformations of dynamical systems must be canonical and are generated 
by physical quantities.

Since the structure of Minkowski's space-time can be derived 
from little more than the most general linear interaction of two 
canonical pairs, this suggests a presentation in which space-time 
(geometry) is derived from dynamics and not vice-versa.
While the second example was based on the Euklidean/Newtonian 
meta-physics of absolute space and time, the priorization has changed 
with the Dirac equation: now the nature of space-time stems from
the structure of the underlying phase space as represented
by the Dirac Clifford algebra. But we do not simply postulate to use 
some Clifford algebra, possibly for reasons of convenience. 
We have shown that the use of Clifford algebras can be motivated 
from Hamiltonian symmetries only\footnote{For more
details see Refs.~\cite{qed_paper,osc_paper}.}, and they receive 
additional and important constraints from these Hamiltonian symmetries,
namely the distinction between Hamiltonian and skew-Hamiltonian (matrix)
elements. 

Lorentz transformations, the Lorentz force and even Maxwell's equations 
are obtained by this type of Hamiltonian deduction.
This raises the question, if one could possibly formulate a similar 
approach, based on larger phase spaces, for a hypothetical world with 
more or less than $3+1$ dimensions. We are not going to discuss this in 
detail, we just mention some restrictions resulting from Hamiltonian
notions\footnote{For more details see Refs.~\cite{qed_paper,osc_paper,uqm_paper}.}.

Neither Newtonian physics nor Einstein's relativity provide any
{\it intrinsic} argument for the dimensionality of space-time. 
In both theories space-time is postulated as if it was one of 
the ten commandments\footnote{See also Stenger~\cite{Stenger}.}. 
Here we want to raise awareness for the fact, that other coherent
narratives indeed do exist: Maybe it is wrong to think that physical
theorizing must presume a specific space-time dimension to begin with. 

Clifford algebras $Cl(p,q)$ with Hamiltonian basis exist only in dimension 
$Cl(N-1,1)$. But it is a math fact called ``Bott-periodicity'' that Clifford 
algebras with real matrix representations exist only for certain dimensions, 
namely with $q=1$ and $p=N-1$ we can only have
\begeq
p-q=N-2=0,\,2\,\rm{ mod }\,8\,.
\label{eq_bottr}
\endeq
This means that irreducible space-times in analogy to the Hamiltonian 
derivation of Minkowski space-time exist only for a subset of (hypothetical)
space-times, namely for $1+1$, $3+1$, $9+1$, $11+1$, $\dots$, $25+1$, $27+1$ 
etc. dimensions\footnote{
Apparently also string theorists found reasons to consider ``space-times''
of the some of these dimensions, namely of $10$ and $26$~\cite{Strings}.}.
We think that the mentioned points are remarkable results, 
which demonstrate how {\it restrictive} Hamiltonian notions actually 
are (see also Fig.~\ref{fig_pascal}).

\section{Why ``Hamiltonian'' Notions, not ``Lagrangian''?}
\label{sec_hamilton}

\begin{quotation}
The power of Lagrangian mechanics has caused generations of students to
wonder why it is necessary or even desirable, to recast mechanics in
Hamiltonian form. The answer [...] is that the Hamiltonian formulation is
a much better basis from which to build more advanced methods. The Hamiltonian
equations have an elegant symmetry that the Lagrangian equations lack~\cite{Johns}.
\end{quotation}

Most common textbooks on classical mechanics follow a ``historical''
approach which usually starts with Newtonian mechanics, continues with
Lagrangian mechanics as a formal generalization of Newton's, then
Hamiltonian mechanics (as a variant of Lagrange's) and eventually 
spend a few words on Hamiltonian-Jacobi theory as yet another formal 
method. But after having solved the standard textbook examples using 
Newtonian and Lagrangian methods, it seems to be an exercise in redundancy 
to re-iterate known solutions of standard problems using yet another method. 
It is hardly possible that students grasp the fundamental differences 
between these theories, which can not be found in the solutions of 
age-old textbook problems but in their respective logical construction.
Taught this way, the students must disentangle on their own that
Hamiltonian mechanics is not at all about mechanics, but has a much
broader scope and applicability, especially when combined with
statistical physics. The concepts of force and inertia, central to
Newtonian physics, are mostly absent in Hamiltonian physics.

Newtonian mechanics has, strictly speaking, been falsified by special
relativity and can therefore not any more be regarded as fundamental to physics.
Though it is possible to force the notation of relativistic mechanics
into a Newtonian form, this can not belie the fact that Newtonian physics
requires fundamental revision in relativity. Not so with Hamiltonian notions.
The Hamiltonian function ${\cal H}(q_i,p_i)$ is different in relativistic
mechanics compared to non-relativistic mechanics, but Hamiltonian
methods do not require a specific ontology to be applicable: Hamiltonians
are by no means intrinsically restricted to the form ${\cal H}(q,p)=T(p)+V(q)$.

So what about Lagrangian mechanics? Some modern textbooks on dynamics drop 
the variational approach completely, mostly due to mathematical 
drawbacks\footnote{See for instance Ref.~\cite{Souriau}, §12.98, page 140.}.
Many bright physicists and mathematicians have understood that the Hamiltonian
form is conceptually superior. Cornelius Lanczos for instance wrote
that\footnote{See page 167 in Ref.~\cite{Lanczos}.}
\begin{quotation}
[Hamilton's] equations are entirely equivalent to the original 
Lagrangian equations and are merely a mathematically new form. 
Yet the new equations are vastly superior to the originals. For
derivatives with respect to $t$ appear only on the left-hand sides
of the equations, since the Hamiltonian function does not contain
any derivatives of $q_i$ or $p_i$ with respect to $t$.
\end{quotation}
He further continues\footnote{See page 194 in Ref.~\cite{Lanczos}.}
\begin{quotation}
A further reason why the Hamiltonian equations are superior
to the Lagrangian equations in their transformation properties
is that the number of variables is doubled.
If at first sight this increase seems more of a loss
than a gain, the procedure of coordinate transformations turns 
the liability into an asset. It is of great advantage that we 
can widen the realm of possible transformations by having a 
larger number of variables at our disposal.

Finally, in Lagrangian mechanics we do not possess any systematic method 
for the simplification of the Lagrangian function.
We may hit on ignorable variables by lucky guesses,
but there is no systematic way of producing them. In Hamiltonian
mechanics a definite method can be devised for the systematic
production of ignorable variables and the simplification of the
Hamiltonian function. This method, which reduces the entire
integration problem to the finding of one fundamental function,
finally, in the generating function of a certain transformation, 
plays a central role in the theory of canonical equations and 
opens wide perspectives [...].
\end{quotation}

But there is another problem, rarely spelled out explicitely, namely that the 
{\it principle of least action}, despite it's admitted elegance and beauty, and it's
widespread use in theoretical physics, is inappropriate to serve as {\it the} 
foundational principle of physics. 
As Pulte reported~\cite{Pulte}, not only Jacobi but many classical physicists and
mathematicians criticized this principle: ``Lagrange’s formulation and (or) 
demonstration of the principle of virtual velocities posed a challenge for a 
number of mathematicians from Fourier (1798), de Prony (1798), Laplace (1799), 
L. Carnot (1803), and Ampere (1806) to Cournot (1829), Gauss (1829), Poisson (1833), 
Ostrogradsky (1835, 1838), and Poinsot (1806, 1838, 1846)''.
According to Pulte, Lagrange's attempt to base mechanics on the principle of virtual 
velocities ``leads inevitably to a conflict with the traditional meaning of axiom as a 
self-evident first proposition, which is neither provable nor in need of a proof.''
Hence if we recall the ``original goal'' of science, namely ``clearing up
mysteries''~\cite{Jaynes89}, then Hamiltonian notions should play a central role.

We have been trained to swallow the ``axioms'' of quantum theory, and
it therefore appears somewhat old-fashioned to demand that axioms should
be {\it self-evident first propositions, which are neither provable
  nor in need of a proof}. Many physisists seem to have lost trust in
such apparently naive positions. Wasn't this exactly Newton's mistake --
to believe that space, time and motion were self-evident?
The fundamental theory of nature, quantum theory, we have been told,
is counter-intuitive and requires even a new non-classical logic. 
Why then should we expect that the theory could be based on axioms in 
the classical sense? But as we could show, quantum theory can be developed
mathematically straight from generalized classical Hamiltonian notions and 
classical arguments of (local) causality. The foundation of this formalism are 
Hamiltonian notions which are so vastly general in their essence, that (when
correctly applied) they cannot fail to describe physical reality.

In it's very core, Hamiltonian methods are based on one simple definition,
namely that, for any stable closed physical system with the dynamical variables
$\psi$, there exists a function ${\cal H}(\psi)=\rm{const}$. This is neither
a postulate nor a principle, it is mostly a definition of the term
``closed physical system''.

The value of any axiomatic and deductive method depends on the 
self-evidence of the axioms, whether this is admitted or not\footnote{The
  attribution of developments to individual scientists is
not always as clear as naming conventions suggest:
``The symplectic formulation of Hamiltonian mechanics can be
retraced (in embryonic form) to the work of Lagrange between 1808 and 1811;
what we today call 'Hamilton’s equations' were in fact written down by
Lagrange who used the letter {\it H} to denote the 'Hamiltonian' to honor
Huygens – not Hamilton, who was still in his early childhood at that time!
It is however undoubtedly Hamilton’s great merit to have recognized the
importance of these equations, and to use them with great efficiency in
the study of planetary motion, and of light propagation''~\cite{Gosson}.
}. We are not the first to criticize the arbitrariness of the
axioms of quantum theory. Fuchs, to pick out a prominent name, wrote about the
``standard axioms'' of quantum mechanics~\cite{Fuchs2002}:
\begin{quotation}
The task is not to make sense of the quantum axioms by heaping more structure, 
more definitions, more science-fiction imagery on top of them, but to throw 
them away wholesale and start afresh. 
\end{quotation}
As our examples illustrate, it seems possible to drop the idea that the
finding of ``deep'' and ``profound'' principles will eventually provide
the corner stones of this theory. The decisive starting point of physical
science is not necessarily a list of counter-intuitive axioms that are
``inscribed on stone tablet''~\cite{Stenger}. The decisive starting point,
in our opinion, are the primary distinctions that the theory is based
upon. 

In the conventional lore Newtonian physics is based on axioms that provide 
some substantial knowledge about the physical world, namely that force and 
acceleration are related by $F=m\cdot a$. 
This equation however represents not a fact about nature (as general
relativity has shown), it is not even a counter-factual assertion about
nature: in the first place it is a definition and a distinction.
It is a definition of what the theory is about. Even if Newton's definitions
would be circular, as occasionally suspected~\cite{Leech}, they nonetheless
provide an account of the conceptional framework to be used. It is a
distinction between quantities that are contingent (initial position and velocity) 
and those that require (and allow for) explanation, namely acceleration. 
As Poincare argued, in many ways Newton's theory contains core elements of 
Hamiltonian mechanics: The initial state of a physical
object is specified by pairs of values for each coordinate, namely by
coordinate position and -velocity. This indirect insight is closely related
to the notion of phase space in Hamiltonian
mechanics\footnote{It is not our intention to question
the value of Newton's (or Heisenberg's) work or to violate the feelings of
those who desire to celebrate the achivements of these giants. But beyond
legitimate hero worship, students should not be mislead to confuse 
a modern understanding of classical physics with point particles and 
Newtonian ``common sense''.}. Classicality can be defined without
Newton's axioms and then it leads, as we have shown, directly to
the pinnacle of modern physics, namely to Dirac's incredible spinor theory.

With the development of thermodynamics, the principle of energy conservation
{\it practically} became a defining principle in physics: no patent
office (and no properly educated physicist) will accept any suggested device 
or theory that is in conflict with energy conservation. A violation of the 
principle of energy conservation is, by definition, ``un-physical''.
But is energy conservation really a self-evident first proposition, which 
is neither provable nor in need of a proof? Why then did it take so long
to discover this principle? We argue here that their is no need to specify
the kind of conserved quantity. It fully suffices to accept that the
whole universe (or whatever fraction of it) could, at least in principle,
be regarded as a closed system (though only {\it in theory}). Then there 
must be a universal scale, i.e. a universal quantity that is conserved 
over all subsystems. This is what ``energy'' means after all, at least 
operationally. If one regards energy conservation from a more general
point of view, it is essentially identical to what is othewise
called ``object permanence'': It says, from a bird's eye view, 
that any ``substantial'' quantity must be preserved in a physical world.

Hence it suffices that Hamiltonian mechanics derives from {\it some} universal
additive conserved quantity; there is no need to introduce the {\it physical} 
notion of energy apriori. The essential distinction that Hamiltonian 
mechanics rests upon, is between dynamical variables and constants of motion, 
or vulgo: between those things that change and those that do not.
Neither can we think of any ``deeper'' physical distinction, nor of any
that is more essential and more trivial. However, a constant can
only be incorporated properly into pure Hamiltonian physics as a constant 
of motion as we shall argue in Sec.~\ref{sec_COMS}. Furthermore, this 
distinction has a corollary: A causal theory requires that change must 
be derivable smoothly from the actual state of affairs.

As we mentioned in the beginning, Hamiltonian mechanics is further profaned
(metaphysically) by the ``theorem due to Lie and Koenigs on the reduction
of any system of ordinary differential 
equations to the Hamiltonian form.''\footnote{See page 275ff in
  Ref.~\cite{Whittaker}, and Ref.~\cite{Santilli}.}. Since any 
classical dynamical theory has the form of a set of ordinary differential 
equations, any classical theory can be reduced to the Hamiltonian form, at 
least {\it locally}. Then Hamiltonian mechanics is little more than a 
``method'' and boils down to the mere possibility to 
describe some physical system by a number $\nu$ of variables 
$\psi=(\psi_1,\psi_2,\dots,\psi_\nu)^T$ that obey some set of ordinary 
differential equations
\begeq
\dot\psi=f(\psi)\,.
\label{eq_dfeq}
\endeq
Then, within the limits of applicability of this theorem, any 
dynamical law can be constructed from a conservation law. These facts
suggests that Hamiltonian mechanics is in itself a
``law without law''\footnote{We shall come back to this in part two.}.

\subsection{Units, Artifacts and Constants of Motion}
\label{sec_COMS}

\begin{quotation}
Metrology truly is the Mother of Science!~\cite{Hall}
\end{quotation}
There is another logical reason to prefer -- at the foundation of physics -- 
a conservation law over Eq.~\ref{eq_dfeq} and 'Hamiltonian' over 'Lagrangian' 
methods. This reason is so basic and simple that it is rarely acknowledged 
at all. Einstein casually raised the issue in a contemplation 
on special relativity~\cite{EinsteinBio}:
\begin{quotation}
It is striking that the theory (except for the four-dimensional space)
introduces two kinds of things, i.e. (1) measuring rods and clocks, (2)
all other things, e.g., the electromagnetic field, the material point, etc.
This, in a certain sense, is inconsistent; strictly speaking, measuring rods
and clocks should emerge as solutions of the basic equations [...], not, as
it were, as theoretically self-sufficient entities.
\end{quotation}
While good introductory textbooks on physics should contain a passage on
weights and measures, most advanced textbooks (and theories) take their 
existence for granted. However, as Bunge remarked,
\begin{quotation}
Being a theoretical concept, the notion of a unit should be elucidated within 
a theory proper. And being a generic notion, i. e. one occurring in every 
branch of quantitative science, its elucidation through a mathematical theory 
is a task for the foundations of science\cite{Bunge}.
\end{quotation}
The short version of our argument is as follows: Weights and measures are,
evidently, not only the practical but also the logical basis of physics as
a quantitative ``exact'' experimental science. If we are supposed to provide 
a theoretical account of weights and measures, for instance of a measuring 
rod of a certain length, then this {\it implies} that we presume further 
{\it underlying} (maybe yet unknown) laws of physics, equations from a
{\it more} fundamental level of reality, from which standards {\it can} 
emerge, at least in principle. The mass of a weight and the length of a rigid 
rod can only be constant, if they are functions of other conserved
quantities that originate on some more fundamental level. We have no other 
idea of how they could physically {\it emerge}, within Hamiltonian theory, 
in any other way than as constants of motion of the underlying microphysical 
dynamical system:
Either one finds a physical system in which a distance or radius (or some
other physical quantity) is a conserved quantity or one derives mathematical 
relationships which allow to express a constant distance as a function of 
other conserved quantities. The precondition for the measurement of a 
physical quantity is the existence of some physical object or artifact for 
which a property of the same physical dimension is a (conserved) constant 
property. But if the precondition to measure some quantity is the availability
of an artifact providing a constant reference standard of the same type and 
dimension, then there must necessarily be one more level of dynamical
quantities -- {\it below} the level of the most fundamental measurable 
quantities. 

Eventually this implies that either there is no fundamental level 
at all, or if there is something like a most basic level, then it consists of 
dynamical variables that can not be directly measured, because by definition
no level {\it below} this most basic level may exist that could provide a 
measurement standard.
This most fundamental level must therefore be represented by variables for
which no reference standard is available. Then they can not be measured
``directly'' and the corresponding variables must remain 
abstract~\cite{qed_paper}: These most fundamental variables are phase
space coordinates, but in an {\it abstract} phase space, in a space
without predefined ontological interpretation.

Hence, also from the standpoint of metrology, we must inevitably
presuppose the validity of conservation laws before we can even start to make
our first measurement: But since {\it some} conservation law has to be presumed 
{\it anyhow}, what motivation remains to ``derive'' them from a 
``least action principle''?

\section{Summary and Conclusions}

The prevalent historically oriented presentation of physical theories 
has drawbacks. Not only does the historical account presented in standard
textbooks often lack historical precision, it is evidently not even intended
to give a correct {\it historical} account, not to speak of an account that would 
be subscribed by professional historians of
science~\cite{McCutchen1972,Whitaker1979a,Whitaker1979b,Kragh1992,Leite2002}.

Furthermore, as Ralston pointed out, {\it the order matters}~\cite{Ralston}.
To place something in an early stage suggests fundamental relevance. 
For instance, since the order of presentation in textbooks on classical 
mechanics usually begins with Newton's axioms, continues with Lagrangian 
mechanics and finally ends up in Hamiltonian mechanics (sometimes followed 
by Hamilton-Jacobi theory), students are lead to conclude that force is 
fundamental and that Hamiltonian notions are but an abstract reformulation 
of Newtonian mechanics. This however is a mis\-re\-pre\-sentation of the logical 
hierarchy. Hamiltonian methods neither require Newton's axioms nor his
space-time metaphysics to achieve meaning and validity. As we have shown,
it is rather the other way around: Newtonian mechanics can be reconstructed
as an approximation of results that can be derived using pure Hamiltonian
methods. The advantage of the latter approach is the possibility to avoid
obscure metaphysical axioms.

Not only did the Hamiltonian methods survive the scientific revolutions
of the 20th century, due to their abstractness and generality one
must conclude that they can be applied to any imaginable physical reality.
As Whittaker taught a century ago, they are but a general mathematical
set of analytical methods that can be applied whenever we consider
dynamical systems that depend on a timelike parameter.
Therefore, we think, it is inappropriate to merely distinguish between
classical physics and quantum physics. 
There is the old ``phenomenological'' classical mechanics (Newton's) and 
the new abstract classical mechanics (Hamilton's). They differ as much as 
the old ``phenomenological'' quantum theory of Bohr and Heisenberg differs 
from the new abstract quantum theory of Schr\"odinger and 
Dirac\footnote{See also Chap. 1-3 in Ref.~\cite{Ralston}.}. 
In both cases there is an old theory that is contaminated with metaphysical 
presuppositions and a new theory in which this extra baggage can be dismissed 
as dead weight~\cite{ODJohns}:
\begin{quotation}
In 1924, Louis de Broglie proposed that particles like electrons also
have wave properties. The new quantum theory by Erwin Schr\"odinger
in 1926 is a wave theory. But the developers of the old quantum theory
had no thought that matter might be wave-like. For them, matter followed
deterministic classical orbits, restricted only by the quantum conditions
imposed on the action variables of the action--angle theory. Given that
major conceptual flaw, it is surprizing that the old quantum theory was
able to explain as much of the experimental physics of the day as it did.
\end{quotation}

Coming back to the question raised in the setup: Apparently one can
find Hamiltonian descriptions with unexpected predictive power because
different levels of physical description are ``vertically'' connected 
by their respective Hamiltonian constraints. 
In the proper Hamiltonian formulation of Dirac's theory no constraints 
from a spatio-temporal level exist and therefore all mathematically 
possible terms, all {\it Hamiltonian} terms, can (and do) have physical 
significance. In this sense, the phase space of Dirac spinors is, from
a logical point of view, more fundamental than Minkowski's space-time:
As we have shown, one can derive the Minkowski-metric from Hamiltonian
notions via Dirac's theory, but the reverse, the introduction 
of the Dirac matrices was always regarded as an ingenious ad-hoc move,
as something that can not be derived but can only be postulated.

It is a historical contingency, namely the historical order in which
quantum theory was developed, that lead quantum physicist to 
believe that imaginary numbers had some new and mysterious 
``non-classical'' significance in quantum mechanics. But if complex
numbers were anything but pairs of real numbers, then one could not
simulate quantum theory with conventional computers, which are known
to be based on integer and floating-point numbers only.
Furthermore, if imaginary numbers had special significance, then there
were $15$ canonical (traceless) generators in Dirac's theory while there
are only ten. The conviction that complex numbers
are, somehow, more fundamental than real numbers, puts forth strange 
blossoms and as of today there is barely any mention of real representations
of anything in theoretical physics. The Hamiltonian account of Dirac's
theory however uncovers an intrinsic redundancy: Dirac's theory does not
require complex wave-functions (though the complex notation might be
easier to handle):
The six skew-Hamiltonian elements, namely the $4$ components of the axial 
vector, the scalar, and the pseudo-scalar are no canonical generators
in real linear Hamiltonian theory, while they would be allowed in a 
complex theory in which matrix transposition is connected with complex
conjugation. Only the use of the real matrices and classical notation
uncovers that the unit imaginary is an artifact from Schr\"odinger's 
theory that does not intrinsically belong to the Dirac equation.

Freeman Dyson complained already in 1962~\cite{Dyson1962}:
\bquo
Probably all these connections would have been clarified
long ago, if quantum physicists had not been hampered 
by a prejudice in favor of complex and against real
numbers.
\equo
And also Dirac had few tendency to be mystified by
the appearance of complex numbers in quantum theory~\cite{Dirac74}:
\bquo
Thus a complex Hilbert vector is not a more general kind of 
quantity than a real one. A real Hilbert space is the more 
elementary concept. A complex Hilbert space should be looked 
upon as a real one in which a certain structure is introduced, 
namely a pairing of the coordinates, each pair being then 
considered as a complex number. Changing the phase factors of 
these complex numbers then provides a special kind of rotation 
in the Hilbert space.
\equo
and he continues:
\bquo
In a structureless real Hilbert space there are no special
linear transformations. All are on the same footing. This is
the most suitable basis for a general mathematical theory.
The existence of special transformations would complicate
the discussion of the fundamental ideas. We shall therefore
deal with a real Hilbert space, where the vectors have real
coordinates.
\equo
However, vice versa, if we are to derive Schr\"odinger's equation
from Dirac's, it is convenient to invent the unit imaginary, simply
to avoid the spinor notation. Hence the unit imaginary entered quantum
mechanics as an artifact of the historical order. 

It has been shown in Ref.~\cite{uqm_paper}, the Dirac current
is the source of the field terms in ${\bf S}$: Maxwell's theory,
the Dirac ``particle'', quantum theory and special relativity can
be developed and presented in one single cast~\cite{qed_paper,osc_paper}. 
Then it might also be possible to explain the statistical properties of 
the ``quantum world'': 
Since the ``particle'' properties are defined by the second moments 
of the phase-space distribution, it is clear why $\vec p=0$ does not imply
$\langle\vec p^2\rangle=0$: In a phase space distribution we may as
well have $\langle p\,q\rangle=0$ {\it and}  $\langle p^2\,q^2\rangle>0$.
To regard this as an intrinsic ``uncertainty'' or ``indeterminacy'' is 
possible but somewhat willful, because fourth order moments do
not necessarily indicate ``uncertainty'' or ``indeterminacy'' of
second order moments. 

We claim that a proper understanding of the mathematical form of
special relativity and QM requires little more than to understand the 
fundamental significance of conserved quantities and Hamiltonian notions.
Then it becomes evident that the math does not describe a particle without
volume but a set of four related quantities (energy and three momentum
components) which are connected in the Dirac theory: They are (linear combinations of)
matrix elements. One can look at them {\it as if} they would represent
a massive point particle. The spin indicates motion, but not {\it spatial} 
motion. The Lorentz transformations are, in the first place, not space-time
coordinate transformations but canonical transformations in spinor space.
They can also be looked upon {\it as if} they would describe the
transformations between ``inertial frames''. 
Many physicists seek for an understanding that goes beyond mere formal 
derivations, for a visualizable setup to begin with. But on the other hand, 
natural philosophy was always driven by the desire to understand the last
principles {\it behind} the apparent phenomena. However, if these principles
are supposed to be found {\it behind} the apparent phenomena, then it would
be odd to expect that they can be obtained directly from the phenomena:
It is no surprize that these principles are not suitable for a spatio-temporal
visualization. But the principles are not ``deep'': In order to serve
as axioms, principles must be (self-) evident, which is the opposite of
deep. ``Deep'' are those consequences that were, due to the simplicity of the
axioms, initially unexpectable. 

So why do we call the Hamiltonian logic ``vertical''? It is vertical insofar
it concerns the logic between two (or more) different levels of description,
namely that
of the dynamical variables $\psi$ of spinorial phase space and that of moments
(observables) of this space, namely ${\bf S}$. Let us reconsider the emergence of 
Eq.~\ref{eq_Heisenberg}: The time derivative of the matrix 
${\bf S}$ on the left side is proportional to elements of ${\bf F}$ 
and to matrix elements of ${\bf S}$ on the right. But the equations of motion
that lead to Eq.~\ref{eq_Heisenberg} stem from one level below, namely from
the equations of motion of $\psi$ (and the {\it same} elements of ${\bf F}$).
This is the vertical connection between the different levels. On the level
of the Dirac spinor, i.e. the supposed fundamental level, the only restriction
of the equations of motion is their Hamiltonian form. On the next level, the
equations of motion

Due to the universality of the Hamiltonian method, it is practically always 
possible to find a Hamiltonian formulation, even if the underlying level is 
either unknown or regarded as metaphysically obscure. However, if one finds
that a level in which {\it all possible} Hamiltonian terms have physical 
significance and none is missing, this is a strong indication that this
respective level is irreducible and hence in this sense {\it fundamental}.
The phase space of Dirac spinors appears to be such a level.

Now the ten Clifford parameters $f_i$ (Eq.~\ref{eq_HamiltonioanMatrix}) of some
Hamiltonian matrix of $sp(4,\mathbb{R})$, also of ${\bf S}$, can be written in 
vector form ${\bf x}=(f_0,f_1,\dots,f_9)^T$ and then Eq.~\ref{eq_Heisenberg}
can be reformulated as~\cite{rdm_paper}:
\begeq
{\bf\dot x}_{i}={\cal F}_{ij}\,{\bf x}_{j}\,.
\endeq
Doing this one may recover the conventional relativistic ``tensor'' notation:
The upper left $4\times 4$ sub-matrix of the $10\times 10$-matrix ${\cal F}$ 
is the electromagnetic field tensor~\cite{rdm_paper}. But the footprint of 
the ``spinorial phase space'', remains on this ``higher'' level, namely the
specific form of ${\cal F}$ and the necessity to use two types of indices, 
i.e. to implement the signature of the Clifford algebra via the use of the 
so-called ``metric tensor''. The matrix ${\bf F}$ has the size $10\times 10$, 
hence many more {\it possible} than actual terms compared to 
the matrix ${\bf F}$.
If an agent (``observer'') has direct access only to (even) moments of $\psi$, 
namely here ${\bf S}$ (or ${\bf x}$, respectively), she is lead to think that 
those variables are fundamental and that the form of ${\cal F}$, reflects some 
fundamental law of nature, while ``in reality'' all that is just statistics
on a lower dimensional space in which no such laws exist. No doubt that such 
an agent might face some unexpected correlations, namely between moments of
the variables in ${\bf x}$.

The usual textbook presentation of physics does less than something to clarify
the vertical logic of Hamiltonian methods that we tried to point out in this 
article; it does not even seem to acknowledge that any original Hamiltonian
logic exists. 

\begin{acknowledgments}
We thank H. R\"ocken for comments and critical remarks.
\LaTeX\ has been used to write this article, XFig 3.2.5 for the figures.
Mathematica\textsuperscript{\textregistered} has originally been used for
parts of the symbolic calculations. 
\end{acknowledgments}

\begin{appendix}

\section{Clifford Algebra in a Nutshell}
\label{sec_CA}

A Clifford algebra $Cl(p,q)$ is generated from $N=p+q$ mutually anti-commuting 
``basis'' elements $\y_\nu$ where $\nu\in[0,\dots,N-1]$, such that
$\y_\nu^2={\bf 1}$ for $\nu\in[0,\dots,p-1]$ and $\y_\nu^2=-{\bf 1}$ for 
$\nu\in[p,\dots,p+q]$. It follows from this definition that the anti-commutators
of the basis elements can be summarized by the so-called ``metric tensor'' $g_{\mu\nu}$:
\begeq
\y_\mu\,\y_\nu+\y_\nu\,\y_\mu=2\,g_{\mu\nu}=2\,\mathrm{Diag}(1,1,\dots,-1,-1,\dots)\,.
\endeq
where $g_{\mu\nu}$ is a diagonal matrix with $p$ diagonal elements equal to 
$+1$ and $q$ diagonal elements equal to $-1$, corresponding to the signature 
of the basis elements $\y_\nu$.

\begin{figure}[h]
\parbox{8.5cm}{
\includegraphics[width=8.5cm,keepaspectratio]{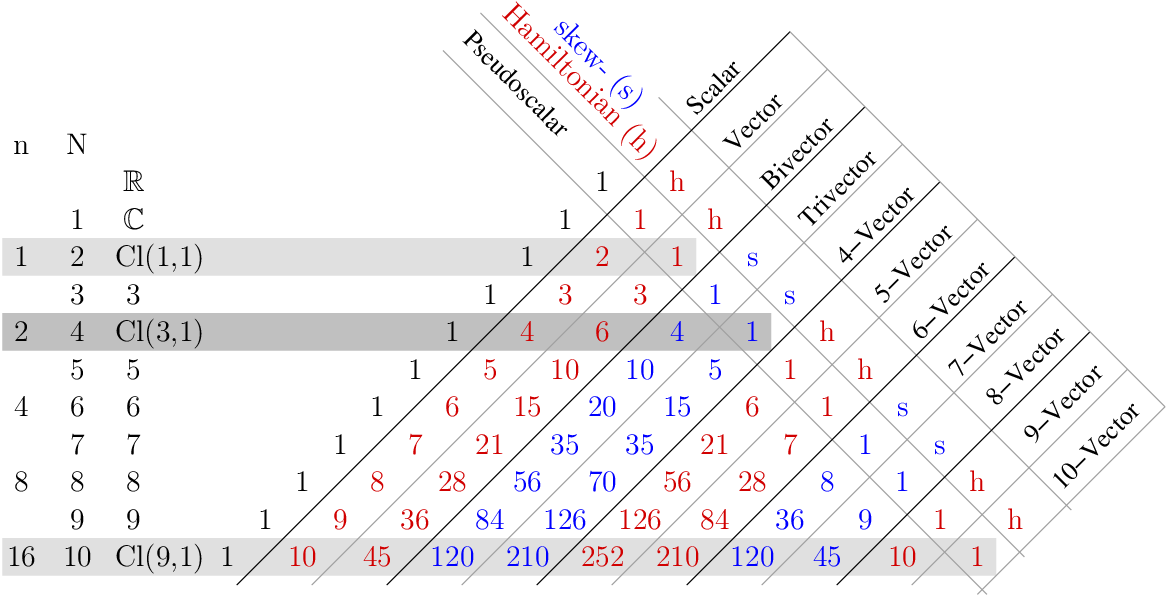}
\caption[Pascal's Triangle]{
Pascal's triangle for Clifford algebras. 
Left: Number of ``spinor components'' $n$, corresponding
dimension $N$ of Clifford algebra and special cases:
$\mathbb{R}=Cl(0,0)$, $\mathbb{C}=Cl(0,1)$.
The rows of the Hamiltonian Clifford algebras are indicated
in gray. Note that a ``Lorentz 4-vector'' in tensor algebra is 
just a 1-vector with 4 components in a Clifford algebra, 
while the ``4-vector'' in the Dirac algebra has only a 
single component called ``pseudoscalar''. 
\label{fig_pascal}
}}
\end{figure}

Only these basic elements $\y_\nu$ are required to generate all other 
elements as (multiple) products of basic elements. 
From combinatorics it is known that a system of $N$ elements allows
for $\left({N\atop k}\right)$ products of $k$ elements and hence
generates a multiplicative group with a total number of
\begeq
\sum\limits_{k=0}^N\,\left({N\atop k}\right)=2^N
\endeq 
elements. The elements are called $k$-vectors, if they are
proportional to products of $k$ basis-elements $\y_\nu$. The product
of all basis elements, the $N$-vector, is called pseudo-scalar.
This means that Clifford algebras are related to Pascal's triangle.
The unit matrix is called {\it scalar} and the product of all basis 
elements is the so-called {\it pseudo-scalar} (see Fig.~\ref{fig_pascal}).
As freely defined mathematical entities, the unit elements $\y_\mu$
do not require any representation beyond a mere symbol and the 
definition given above. Framed just mathematically one may 
define and analyze CAs with an arbitrary number of dimension 
and any signature. This is certainly an interesting (and active) 
field of research in its own right, but it is not of specific 
interest here\footnote{
Standard textbooks are, for instance, Ref.~\cite{Lounesto,Porteous}.
}. 
As physicists we are most often interested in (matrix) representations 
of CAs. But matrix elements may, according to the prevalent reading,
either be real or complex numbers and even quaternions. Regarded this
way, also matrices may be matrix elements and since the complex 
numbers and quaternions\footnote{
The complex numbers $\mathbb{C}$ and the quaternions $\mathbb{H}$
have no irreducible representation by real matrices, but they 
are Clifford algebras, $Cl(0,1)$ and $Cl(0,2)$, respectively.
$\mathbb{C}$ can be represented by a subalgebra of the real 
Pauli algebra and $\mathbb{H}$ can be represented by two different
subalgebras of $Cl(3,1)$.
}
are in themselves representations of Clifford algebras, one may also 
use (why not?) Clifford algebraic elements within matrices. 
Yet again, representation theory is an interesting (and active) 
field of research in its own right, but here we are only interested 
in CAs insofar as they allow for the analysis of classical Hamiltonian 
symmetries\footnote{
See, for instance, Ref.~\cite{FultonHarris}.}. This suggests a 
restriction to real matrices, but this is not really a reduction of
the possibilities: Any Clifford algebra can, in some way, be 
represented by real matrices, because, as we just mentioned,
also the complex numbers and the quaternions are Clifford algebras
in themselves and have real matrix representations. 

The complex numbers, for instance, require a single unit 
element $i$ with $i^2=-1$. We could also say, it consists
only of the SUM $\y_0$ and the unit matrix ${\bf 1}$.
This is the Clifford algebra $Cl(0,1)$. A representation
by real matrices is possible, but ``incomplete'' insofar
as the required matrices allow for a larger algebra than
the complex numbers: Regarded this way, the complex numbers 
are a sub-algebra of the real Pauli algebra.

The next step would be an algebra with two basis elements,
say the Pauli matrices $\eta_0$ and $\eta_1$ with $\eta_0^2=-{\bf 1}$
and $\eta_1^2={\bf 1}$. The only other element (besides the
neutral element, i.e. unit matrix), according to
Pascal's triangle, then is $\eta_0\,\eta_1$, which then is
both, the only existing bi-vector and the pseudo-scalar (see
Fig.~\ref{fig_pascal}). 

From a conceptional point of view, representations based
on complex numbers and quaternions are ``tricky'' because they
use structures inside structures. As we have shown in Sec.~\ref{sec_ex3},
the Hamiltonian way to regard CAs is based on the idea to charge
numbers with structural meaning. But it is somewhat pointless to 
charge structures with structural meaning. Therefore, from a puristic 
Hamiltonian point of view, only those Clifford algebras which
have an irreducible real matrix representation are of primary
concern.

Now let's consider the algebra $Cl(3,0)$ which consists
of $3$ basis elements, ${\bf e}_1$, ${\bf e}_3$ and ${\bf e}_3$
and regard the ``vectors'' ${\bf x}=x\,{\bf e}_1+y\,{\bf e}_2+z\,{\bf e}_3$
and ${\bf p}=p_x\,{\bf e}_1+p_y\,{\bf e}_2+p_z\,{\bf e}_3$
just as we would write vectors in classical vector algebra.
Let us have a look at the respective (anti-) commutative products
of two such vectors (compare to Eq.~\ref{eq_mag}):
{\small\myarray{
{\bf x}\,{\bf p}&=(x\,p_x+y\,p_y+z\,p_z)\,{\bf 1}\\
&+(y\,p_z-z\,p_y)\,{\bf e}_2\,{\bf e}_3\\
&+(z\,p_x-x\,p_z)\,{\bf e}_3\,{\bf e}_1\\
&+(x\,p_y-y\,p_x)\,{\bf e}_1\,{\bf e}_2\,,
}}
\vspace{5mm}

The result contains, firstly, a scalar component equal to the scalar 
product of classical vector algebra, and secondly the vector (``cross'') 
product ${\bf x}\times{\bf p}$ appearing in the coefficients of the bi-vectors.

Hence we find the ``meaning'' of commutative (outer) and anti-commutative
(inner) products 
\myarray{
  {\bf x}\cdot{\bf p}&=({\bf x}\,{\bf p}+{\bf p}\,{\bf x})/2\\
  {\bf x}\wedge{\bf p}&=({\bf x}\,{\bf p}-{\bf p}\,{\bf x})/2\\
}
This gives a first glimpse of why Clifford algebras are said to have 
{\it geometric} content. For a detailed discussion of the general 
Lorentz covariance as represented by $Cl(3,1)$ see Ref.~\cite{lt_paper}.

\section{The General Hamiltonian Matrix}
\label{sec_GHM}

In the chosen matrix representation, the particle's 
matrix is (compare Eq.~\ref{eq_Fvec}, Eq.~\ref{eq_Fvec1}):
{\small\begeq
{\bf P}=\bmtx{cccc}
-P_z&{\cal E}-P_x&0&P_y\\
-{\cal E}-P_x&P_z&P_y&0\\
0&P_y&-P_z&{\cal E}+P_x\\
P_y&0&-{\cal E}+P_x&P_z\\
\emtx
\label{eq_Pmtx}
\endeq}
and that of the electromagnetic fields is (Eq.~\ref{eq_Fbivec}):
{\small\begeq
{\bf F}=\bmtx{cccc}
-E_x&E_z+B_y&E_y-B_z&B_x\\
E_z-B_y&E_x&-B_x&-E_y-B_z\\
E_y+B_z&B_x&E_x&E_z-B_y\\
-B_x&-E_y+B_z&E_z+B_y&-E_x\\
\emtx
\label{eq_Fmtx}
\endeq}

\end{appendix}

\bibliography{vlh_paper_part1.bib}{}
\bibliographystyle{unsrt}

\end{document}